\documentclass{HREstimation}

\usepackage{amsmath,natbib}

\RequirePackage{hypernat}

\usepackage[pdftex]{graphicx}

\usepackage{amssymb}
\usepackage{ifthen}
\usepackage{enumitem}
\usepackage{grffile} 

\newtheorem{satz}{Satz}[section]

\newtheorem{prop}[satz]{Proposition}
\newtheorem{theo}[satz]{Theorem}

\newtheorem{rem}[satz]{Remark}

\newcommand{\R}{\mathbb R}
\newcommand{\E}{\mathbb E}

\newcommand{\N}{\mathbb N}

\renewcommand{\P}{\mathbb P}

\newcommand{\0}{\mathbf 0}
\newcommand{\1}{\mathbf 1}

\newcommand{\Xbf}{\mathbf X}
\newcommand{\Ybf}{\mathbf Y}

\newcommand{\sbf}{\mathbf s}

\newcommand{\xbf}{\mathbf x}
\newcommand{\Xcal}{\mathcal X}

\DeclareMathOperator{\Var}{Var}

\DeclareMathOperator{\diag}{diag}

\DeclareMathOperator{\mean}{mean}
\DeclareMathOperator{\mado}{mado}
\DeclareMathOperator{\HRMLE}{HRMLE}
\DeclareMathOperator{\MLEa}{MLE}
\DeclareMathOperator{\MLEb}{MLE2}
\DeclareMathOperator{\SPEC}{SPEC}
\DeclareMathOperator{\PROJ}{PROJ}

\author{Sebastian Engelke}
\address{Institut f\"ur Mathematische Stochastik,
  Georg-August-Universit\"at G\"ottingen, 
  Goldschmidtstr. 7, 37077 G\"ottingen, 
  Germany.}
\email{sengelk@uni-goettingen.de}

\author{Alexander Malinowski}
\address{Institut f\"ur Mathematische Stochastik,
  Georg-August-Universit\"at G\"ottingen, 
  Goldschmidtstr. 7, 37077 G\"ottingen, 
  Germany.}
\email{malinows@math.uni-goettingen.de}

\author{Zakhar Kabluchko}
\address{Institut f\"ur Stochastik, 
  Universit\"at Ulm, 
  Helmholtzstr. 18, 89069 Ulm,
  Germany.}
\email{zakhar.kabluchko@uni-ulm.de}

\author[S. Engelke, A. Malinowski, Z. Kabluchko, M. Schlather]{Martin Schlather}
\address{Institut f\"ur Mathematik,
  Universit\"at Mannheim, 
  A5, 6, 68131 Mannheim,
  Germany.}
\email{schlather@math.uni-mannheim.de}

\title[Estimation of H\"usler-Reiss distributions and Brown-Resnick processes]{Estimation of H\"usler-Reiss distributions\newline and Brown-Resnick processes}

\begin{document}

\begin{abstract}
Estimation of extreme-value parameters from observations in the
max-domain of attraction (MDA) of a multivariate max-stable
distribution commonly uses aggregated data such as block maxima. Since
we expect that additional information is contained in the
non-aggregated, single ``large'' observations, we introduce a new
approach of inference based on a multivariate peaks-over-threshold
method. We show that for any process in the MDA of the
frequently used H\"usler-Reiss model or its spatial extension, the
Brown-Resnick process, suitably defined conditional increments
asymptotically follow a multivariate Gaussian distribution. This leads
to computationally efficient estimates of the H\"usler-Reiss parameter
matrix. Further, the results enable parametric inference for 
Brown-Resnick processes.\\ A simulation study compares the
performance of the new estimators to other commonly used methods. As
an application, we fit a non-isotropic Brown-Resnick process to the
extremes of 12 year data of daily wind speed measurements.
\end{abstract}

\section{Introduction}

Univariate extreme value theory is concerned with the limits of
linearly normalized maxima of i.i.d.\ observations, namely the
max-stable distributions (cf.\ \citet{deh2006a}). Statistical
inference of the parameters is well-developed and usually based on one
of the following two approaches.  Maximum likelihood
estimation is applied to blockwise maxima of the original data, where a typical block size in
environmental applications is one year. On the other hand, the
\emph{peaks-over-threshold} (POT) method fits a suitable Poisson point
process to all data that exceed a certain high threshold and thus
follow approximately a generalized Pareto distribution
(cf.\ \citet{dav1990}). The advantage of the latter approach is that it
avoids discarding extreme values within the blocks that are below the
maximum but nevertheless contain information on the parameters.\\ When
interested in the joint extreme behavior of multivariate
quantities, there are different possibilities of ordering the
data, though, the most common procedure is taking componentwise
maxima. In multivariate extreme value theory, a random process
$\{\xi(t):t\in T\}$ with some index set $T$ is called
\emph{max-stable}, if there exists a sequence $(\eta_i)_{i\in\N}$ of
independent copies of a process $\{\eta(t):t\in T\}$ and functions
$c_n(t) > 0$, $b_n(t)\in \R$, $n\in\N$, such that the convergence
\begin{align}
  \label{xi_conv}
  \xi(t) = \lim_{n\to\infty} c_n(t)
  \left(\max_{i=1}^n \eta_i(t) - b_n(t)\right), 
  \quad t\in T,
\end{align}
holds in the sense of finite dimensional distributions. In this case,
the process $\eta$ is said to be in the max-domain of attraction (MDA)
of $\xi$. Typically, $T$ is a finite set or $T = \R^d, d\in\N$, for
the multivariate or the spatial case, respectively. Both theory and
inference are considerably more demanding than in the univariate
framework due to the fact that no finite-dimensional parametric model
captures every possible dependence structure of a multivariate
max-stable distribution (cf.\ \citet{res2008}).  Similarly to the
univariate case, a standard approach for parameter estimation of the
max-stable process $\xi$ from data in its MDA is via componentwise
block maxima, which ignores much of the information contained in the
original data.  Moreover, even if the exact max-stable process is
available, maximum likelihood (ML) estimation is problematic since
typically only the bivariate densities of max-stable distributions are
known in closed form. Composite likelihood (CL) approaches are common
tools to avoid this difficulty (cf.\ \citet{pad2010},
\citet{dav2012}).\\ Only recently, multivariate POT methods have
attracted increased attention. In contrast to the univariate case, the
definition of exceedances over a certain threshold is ambiguous. For
instance, \citet{roo2006} define a multivariate generalized Pareto
distribution (MGPD) as the limit distribution of some multivariate
random vector in the MDA of a max-stable distribution, conditional on
the event that at least one of the components is large. A simulation study in
\citet{bac2012} shows, that these MGPD perform well in many
situations, yet, again only bivariate densities in a CL framework are
used since multivariate densities are unknown. Alternatively,
exceedances can be defined as the event that the norm of the random
vector is large, giving rise to the spectral measure 
(cf.\ \citet{col1991}). \citet{eng2012} have recently proposed to 
condition a fixed component on exceeding a high threshold, which enables
new methods of inference for processes that admit a certain incremental or a mixed moving maxima
representation.

\medskip

With regard to practical application such as modeling extreme wind
speed or precipitation data, max-stable models need to find a
compromise between flexibility and tractability. There are several
parametric families of multivariate extreme-value distributions (see
\citet{kot2000}) and only few max-stable models in the spatial domain
(cf.\ \citet{smi1990, sch2002, deh2006}). 
For most of them, statistical inference is
difficult and time-intensive. Furthermore, except for the
max-stable process $\xi$ in \eqref{xi_conv} itself, usually no further
processes $\eta$ in the MDA of attraction of $\xi$ are known and thus,
it lacks a theoretical connection between modeling the daily
processes $\eta$ and modeling the extremal process $\xi$.\\ In many
applications such as geostatistics it is natural to assume that the
data is normally distributed. Under this assumption, the only
possible non-trivial limit for extreme observations is the $d$-variate
H\"usler-Reiss distribution (cf.\ \citet{hue1989,kab2011}). In fact,
\citet{has2006} and \citet{has2012} show that also other distributions
are attracted by the H\"usler-Reiss distribution. Hence, we can expect
good fits of this model if the daily data is close to normality.
Recently, it has been shown that the class of Brown-Resnick processes
(\citet{bro1977,kab2009}) constitutes the spatial analog of the
H\"usler-Reiss distributions since the latter occur as finite-dimensional
marginals of the Brown-Resnick process. The research on both theoretical
properties (cf.\ \citet{oes2012,dom2011} for simulation methods) and
practical applications (e.g., \citet{dav2012b}) of these processes is
actively ongoing at present.  Statistical inference, however, was so
far limited to the CL methods based on bivariate densities.

\medskip

In this paper, we propose new estimation methods based on a POT
approach for data in the MDA of H\"usler-Reiss distributions and
Brown-Resnick processes. Similarly to \citet{eng2012}, we consider
\emph{extremal increments}, i.e., increments of the data with respect
to a fixed component, conditional on the event, that this component is
large.  The great advantage of this approach is the fact that the
extremal increments turn out to be multivariate Gaussian
distributed. This enables, for instance, ML estimation with the full
multivariate density function as well as parameter estimation based on
functionals of the Gaussian distribution. Moreover, the concept of
extremal increments as well as estimators derived from spectral
densities are shown to be suitable tools for fitting a Brown-Resnick
process based on a parametric family of variograms.

The remainder of the paper is organized as follows. Section
\ref{methods} comprises the definitions and some general properties of
H\"usler-Reiss distributions and Brown-Resnick processes. In Section
\ref{sec_estimation}, we provide a result on weak convergence of
suitably transformed and conditioned variables in the MDA of the
H\"usler-Reiss distribution, which is the basis for our estimation
methods. It is used to derive the specific asymptotic distribution for
extremal increments (Section \ref{sec_theory}) and for conditioning in
the spectral sense (Section \ref{spec_inference}). In both cases,
non-parametric estimation as well as parametric fitting of
Brown-Resnick processes are considered. A simulation study is
presented in Section \ref{sim_study}, which compares the performance
of the different estimators from the preceding section. As an
application, in Section~\ref{wind_data} we analyze daily wind speed
data from the Netherlands and use our new methods of inference to
model spatial extreme events.  Proofs of the theoretical results can
be found in the Appendix.

\section{H\"usler-Reiss distributions and Brown-Resnick processes}
\label{methods}

In this section we briefly review some details on H\"usler-Reiss
distributions and Brown-Resnick processes and define \emph{extremal
  coefficient functions} as a dependence measure for max-stable
processes.
\subsection{H\"usler-Reiss distributions}

The multivariate H\"usler-Reiss distribution was introduced
in \citet{hue1989} as the limit of suitably normalized Gaussian random
vectors. Suppose that the correlation matrix
$\Sigma_n$ in the $n$-th row of a triangular array of $(k+1)$-variate, zero-mean, unit-variance Gaussian
distributions satisfies
\begin{align}
  \label{corr_conv}
  \Lambda = \lim_{n\to\infty} \log (n) (\mathbf{1\cdot 1}^\top - \Sigma_n) 
  \in \mathcal{D}, 
\end{align}
where $\mathbf{1} = (1,\dots,1)^\top\in\R^{k+1}$ and $\mathcal{D}\subset [0,\infty)^{(k+1)\times(k+1)}$ denotes the space of symmetric, strictly conditionally negative definite
matrices 
\begin{align*}
  \mathcal{D} = &\Bigg\{ \left(a_{i,j}\right)_{0\leq i,j\leq k} = A
  \in [0,\infty)^{(k+1)\times(k+1)}: \mathbf{x}^\top A\mathbf{x} < 0
    \text{ for all }\mathbf{x}\in\R^{k+1}\setminus \{\0\} \text{ s.t.\ }
    \\ &\qquad\sum_{i=0}^{k} x_i = 0, \, a_{i,j} = a_{j,i}, a_{i,i} = 0
    \text{ for all } 0\leq i,j\leq k \Bigg\}.
\end{align*}
 Then the normalized
row-wise maxima converge to the $(k+1)$-variate H\"usler-Reiss
distribution which is completely characterized by the matrix
$\Lambda$. Note that $(\mathbf{1\cdot 1}^\top - \Sigma_n)$ automatically lies in $\mathcal{D}$ if $\Sigma_n$ is non-degenerate, $n\in\N$.
 For any matrix $\Lambda = \left( \lambda^2_{i,j}\right)_{0\leq
   i,j \leq k}\in \mathcal{D}$, define a family of positive definite matrices by
\begin{align*}
  \Psi_{l,\mathbf{m}} (\Lambda) = 2\left(\lambda_{m_i,m_0}^2 +
  \lambda_{m_j,m_0}^2 - \lambda_{m_i,m_j}^2 \right)_{1\leq i,j \leq
    l},
\end{align*}
 where $l$ runs over $1, \dots,  k$ and $\mathbf{m} = (m_0,\dots, m_l)$ with $0\leq
m_0 < \dots < m_l\leq k$. The distribution function of the
$(k+1)$-dimensional H\"usler-Reiss distribution with standard Gumbel
margins is then given by
\begin{align}
  \label{HRdistr}
  H_\Lambda (\mathbf{x}) = \exp\left\{\sum_{l=0}^{k} (-1)^{l+1} \sum_{ \mathbf{m}:
    0\leq m_0 < \dots < m_l\leq k } h_{l,\mathbf{m},\Lambda}(x_{m_1},\dots,
  x_{m_l}) \right\},\quad \xbf \in \R^{k+1},
\end{align}
where
\begin{align*}
  h_{l,\mathbf{m},\Lambda}(y_0,\dots,y_l) = \int_{y_0}^\infty S \left\{
  \left( y_i - z + 2\lambda^2_{m_i,m_0} \right)_{i=1,\dots,l} |
  \Psi_{l,\mathbf{m}}(\Lambda) \right\} e^{-z}\, dz,
\end{align*}
for $1\leq l\leq k$ and $h_{0,\mathbf{m},\Lambda}(y) = \exp(-y)$ for $\mathbf{m}\in\{0,\dots,k\}$.
Furthermore, for $q\in\N$ and $\Psi\in \R^{q\times q}$ positive definite, 
$S(\,\cdot\, | \Psi)$ denotes the so-called survivor
function of a $q$-dimensional normal random vector with mean vector
$\mathbf{0}$ and covariance matrix $\Psi$, i.e., if $\mathbf{Y}\sim
N(\mathbf{0},\Psi)$ and $\mathbf{x}\in\R^q$, then $S(\mathbf{x}|\Psi)
= \P \left( Y_1 > x_1, \dots, Y_q > x_q \right)$.
In the bivariate case, the distribution function \eqref{HRdistr}
simplifies to
\begin{align}
  \label{HRbiv}
  H_\Lambda(x, y) 
  = \exp\left\{-e^{-x}\Phi\left(\lambda + \frac{y-x}{2\lambda}\right)
              -e^{-y}\Phi\left(\lambda + \frac{x-y}{2\lambda}\right) 
              \right\}, \quad x, y\in\R,
\end{align} 
where $\lambda = \lambda_{0,1}\in[0,\infty]$ parametrizes between
independence and complete dependence for $\lambda = \infty$ and $\lambda = 0$,
respectively.\\
Note that the class of H\"usler-Reiss distributions is closed in the 
sense that the lower-dimensional margins of $H_\Lambda$ are again
H\"usler-Reiss distributed with parameter matrix consisting of
the respective entries in $\Lambda$. Consequently, the distribution 
of the bivariate sub-vector of the $i$-th and $j$-th component only depends on the
parameter $\lambda_{i,j}$. Thus, one can modify this parameter
(subject to the restriction $\Lambda\in\mathcal{D}$) without affecting
the other components. This flexibility was demanded in \citet{coo2010} as
a desirable property of multivariate extreme value models that most
models do not possess, unfortunately.

\begin{rem}
  The $k$-variate H\"usler-Reiss distribution is usually given by its 
  distribution function $H_\Lambda$. The density for $k\geq 3$ is rather
  complicated and involves multivariate integration. Hence,
  for maximum likelihood estimation based on block maxima,
  only the bivariate or sometimes the trivariate (cf.\ \citet{gen2011})
  densities are used in the framework of a composite likelihood approach.
\end{rem}

\subsection{Brown-Resnick processes}
\label{sec_BR}
For $T=\R^d$, $d\geq 1$, let $\{Y(t): t\in T\}$ be a centered Gaussian
process with stationary increments. Further, let $\gamma(t) = \E(Y(t)
- Y(0))^2$ and $\sigma^2(t) = \E(Y(t))^2$ be the variogram and the
variance of $Y$, $t\in\R^d$, respectively. Then, for a Poisson point process
$\sum_{i\in\N} \delta_{U_i}$ on $\R$ with intensity $e^{-u}du$ 
and  i.i.d.\ copies
$Y_i\sim Y$, $i\in\N$, the process
\begin{align}
  \label{BR_proc}
  \xi(t) = \max_{i\in\N} U_i + Y_i(t) - \sigma^2(t) / 2, \quad t\in\R^d,
\end{align}
is max-stable, stationary and its distribution only depends on the
variogram $\gamma$. For the special case where $Y$ is a Brownian
motion, the process~$\xi$ was already introduced by
\citet{bro1977}. Its generalization in \eqref{BR_proc} is called
\emph{Brown-Resnick process associated to the variogram $\gamma$}
(\citet{kab2009}). Since any conditionally negative definite function
can be used as variogram, Brown-Resnick processes constitute an
extremely flexible class of max-stable random fields.  Moreover, the
subclass associated to the family of fractal variograms
$\gamma_{\alpha,s}(\cdot) = \| \cdot / s \|^\alpha, \alpha\in (0,2],
  s\in (0,\infty),$ arises as limits of pointwise maxima of suitably
  rescaled and normalized, independent, \emph{stationary} and
  isotropic Gaussian random fields (cf.\ \citet{kab2009}). Here
  $\|\cdot\|$ denotes the Euclidean norm. The model by
  \citet{smi1990} is another frequently used special case of
  Brown-Resnick processes, which corresponds to the class of
  variograms $\gamma(h) = \| h \Sigma^{-1} h \|$, for $h\in\R^d$ and
  an arbitrary covariance matrix $\Sigma\in\R^{d\times d}$.\\ We
  remark that the finite-dimensional marginal distribution at
  locations $t_0, \ldots, t_k\in\R^d$ of a Brown-Resnick process is
  the H\"usler-Reiss distribution $H_\Lambda$ with $\Lambda =
  (\gamma(t_i-t_j)/4)_{0\leq i,j\leq k}$.

\subsection{Extremal coefficient function}
Since, in general, covariances do not exist for extreme value
distributed random vectors, other measures of dependence are
usually considered, one of which being the \emph{extremal coefficient
  $\theta$}.  For a bivariate max-stable random vector $(X_1, X_2)$
with identically distributed margins, $\theta\in[1,2]$ is determined by
\begin{align*}
  \P(X_1\leq u, X_2 \leq u) = \P(X_1\leq u)^\theta, 
\end{align*}
for some (and hence all) $u\in\R$. The quantity $\theta$ measures the
degree of tail dependence with limit cases $\theta=1$ and $\theta = 2$
corresponding to complete dependence and complete independence,
respectively.  For a stationary, max-stable process $\xi$ on $\R^d$,
the \emph{extremal coefficient function} $\theta(h)$ is defined as the
extremal coefficient of $(\xi(0), \xi(h))$, for $h\in\R^d$ (\citet{sch2003}).\\ For the
bivariate H\"usler-Reiss distribution \eqref{HRbiv} we have
$H_\Lambda(u, u)=\exp\left(-2\Phi(\lambda)e^{-u}\right)$ and thus, the
extremal coefficient equals $\theta=2\Phi(\lambda)$. Hence, for
H\"usler-Reiss distributions, there is a one-to-one correspondence between
the parameter $\lambda\in[0, \infty]$ and the set of extremal coefficients. 
Similarly, the extremal coefficient function of the Brown-Resnick process 
in \eqref{BR_proc} is given by 
\begin{align}
  \label{ECF}
  \theta(h) = 2 \Phi( \sqrt{\gamma(h)} / 2 ), \qquad h\in\R^d.
\end{align}
Since there are model-independent estimators for the extremal coefficient
function, e.g., the madogram in \citet{coo2006}, it is a common
tool for model checking.

\section{Estimation} 
\label{sec_estimation}
In this section, we propose new estimators for the parameter matrix
$\Lambda$ of the H\"usler-Reiss distribution and use them to fit
Brown-Resnick processes based on a parametric family of variograms.
We will consider both estimation based on extremal increments and
estimation in the spectral domain.\\
Suppose that $\mathbf{X}_{i} =
(X_i^{(0)},\dots,X_i^{(k)})$, $i=1,\ldots,n$, are independent copies
of a random vector $\Xbf\in\R^{k+1}$ in the MDA of the
H\"usler-Reiss distribution $H_\Lambda$ with some parameter matrix $\Lambda = (
\lambda^2_{i,j})_{0\leq i,j \leq k} \in\mathcal{D}$.  Recall that
$H_\Lambda$ has standard Gumbel margins. Without loss of generality,
we assume that $\Xbf$ has standard exponential margins. Otherwise we
could consider $(U_0(X_i^{(0)}),\dots,U_k(X_i^{(k)}))$, where $U_i =
-\log(1-F_i)$, and $F_i$ is the cumulative distribution function of
the $i$-th marginal of $\Xbf$ (cf.\ Prop.\ 5.15 in \citet{res2008}).
In the sequel, we denote by $\tilde{\Xbf}_{n} = \Xbf - \log n$ and
$\tilde{\Xbf}_{i,n} = \Xbf_i - \log n$ the rescaled data such that the
empirical point process $\Pi_n = \sum_{i=1}^n
\delta_{\tilde{\Xbf}_{i,n}}$ converges in distribution to a Poisson
point process $\Pi$ on $E = [-\infty,\infty)^{k+1}\setminus
  \{-\boldsymbol{\infty}\}$ with intensity measure
  $\mu([-\boldsymbol{\infty},\mathbf{x}]^C) = -\log
  H_\Lambda(\mathbf{x})$ (Prop. 3.21 in \citet{res2008}), as
  $n\to\infty$. Based on this convergence of point processes, the
  following theorem provides the conditional distribution of those
  data which are extreme in some sense.

\begin{theo}
  \label{thm_conv}
  For $m\in\N$ and a metric space $S$, let $g:\R^{k+1} \to S$ be a measurable
  transformation of the data and assume that it satisfies the
  invariance property $g(x + a\cdot\mathbf{1}) = g(x)$ for any
  $a\in\R$ and $\mathbf{1} = (1,\dots,1)\in\R^{k+1}$. Further, let
  $u(n)>0$, $n\in\N$, be a sequence of real numbers such that
  $\lim_{n\to\infty} u(n) / n = 0$.  Then, for all Borel sets
  $B\in\mathcal{B}(S)$ and $A\in\mathcal{B}(E)$ bounded away from 
  $-\boldsymbol{\infty}$,
  \begin{align}
    \label{conv_assump}
    \lim_{n\to\infty}\P\left\{ g(\tilde{\Xbf}_n) \in B \,\big|\, 
    \tilde{\Xbf}_n \in A - \log u(n) \right\} 
    = Q_{g,A}(B),
  \end{align}
  for some probability measure $Q_{g,A}$ on $S$.
\end{theo}

\begin{rem}
  Note that due to the invariance property of $g$, the transformed 
  data is independent of the rescaling, i.e. $g(\tilde{\Xbf}_{i,n}) = g(\Xbf_{i})$,
  for all $i=1,\dots,n$, $n\in\N$.
  
  In the above theorem, $u(n)$ only has to satisfy 
  $u(n)/n\to 0$, as $n$ tends to $\infty$. However, for practical
  applications it is advisable to choose $u(n)$ in such a manner
  that also $\lim_{n\to\infty} u(n) = \infty$, since this ensures 
  that the cardinality of the index set of extremal observations 
  \begin{align}
    \label{index_set}
    I_A= \bigl\{i\in\{1, \ldots, n\} : 
    \tilde{\Xbf}_{i,n} \in A - \log u(n)\bigr\},
  \end{align}
  tends to $\infty$  as $n\to\infty$, almost surely.
\end{rem}

Theorem \ref{thm_conv} implies that for all extreme events, the transformed
data $\{g(\Xbf_{i}): i\in I_A\}$ approximately follow the
distribution $Q_{g,A}$.  Clearly, $Q_{g,A}$ depends on the choices for
$g$ and $A$ and in the subsequent
sections we encounter different possibilities for which 
the limit \eqref{conv_assump} can be computed
explicitly. Furthermore, if $g$ and $A$ are chosen suitably, the
distribution $Q_{g,A}$ will still contain all information on the parameter
matrix $\Lambda$. Our estimators will therefore be based on the set of
transformed data $\{g(\Xbf_{i}): i\in I_A\}$ and the knowledge of
their asymptotic distribution $Q_{g,A}$.  For instance, a maximum
likelihood approach can be applied using the fact that $\Pi_n$
converges to $\Pi$. If, for a particular realization of the
$\Xbf_{i}$, $I_A=\{i_1, \ldots, i_N\}$ for some $N\leq n$, $i_1,
\ldots, i_N\in\{1,\ldots, n\}$, and $g(\Xbf_{i}) =
\mathbf{s}_{i}$, $i=1,\dots,n$, a canonical approach is to maximize
the likelihood
\begin{align*}
  L_{g,A}(\Lambda; \sbf_1,\ldots, \sbf_n)&=\P\left\{|I_A|=N,\, 
  g(\Xbf_{i_j})\in d \sbf_{i_j},\, j=1,\ldots,N\right\}\\
  &=\P\left(|I_A|=N\right)
  \prod_{j=1}^N\P\left\{g(\Xbf) \in d \sbf_{i_j} \,|\, \tilde{\Xbf}_n \in A - \log u(n)\right\}.
\end{align*}
With the Poisson approximation
$\sum_{i=1}^n \1\{\tilde{\Xbf}_{i,n}\in A - \log u(n)\} \approx \text{Pois}\{\mu(A - \log u(n))\}$
and the convergence \eqref{conv_assump} we obtain
\begin{align}
\label{gen_likelihood}
L_{g,A}(\Lambda; \sbf_1,\ldots, \sbf_n)
   \approx  \exp\{-\mu(A - \log u(n))\}\frac{\mu(A - \log u(n))^N}{N!} \prod_{j=1}^N
   Q_{g,A}(d\sbf_{i_j}).
\end{align}
If the ML approach is unfeasible, estimation of $\Lambda$ can also be based on other
suitably chosen functionals of the conditional distribution of~$g(\Xbf)$, for instance
on the variance of~$g(\Xbf)$.

\subsection{Inference based on extremal increments}
\label{sec_theory}
In this subsection, we apply Theorem \ref{thm_conv} with $g$ mapping the data
to its increments w.r.t.\ a fixed index, i.e., $g:\R^{k+1} \to \R^k$, $\xbf \mapsto \Delta \xbf =
(x^{(1)} - x^{(0)} ,\dots, x^{(k)} - x^{(0)})$. In particular, $g$ satisfies the invariance
property $g(x + a\cdot\mathbf{1}) = g(x)$ for any $a\in\R$. Consequently, our estimators
are based on the incremental distribution of those data which are extreme in
the sense specified by the set $A$. 
The following theorem provides the limiting distribution $Q_{g,A}$ for two particular choices
of $A$, namely $A_1 = (0,\infty) \times \R^k$ and $A_2 = [-\boldsymbol{\infty},\0]^C$.

\begin{theo}
  \label{thm_HR}
  Let $\mathbf{X}$ be in the MDA of $H_\Lambda$ with some
  $\Lambda\in\mathcal{D}$, and suppose that the sequence $u(n)$ is
  chosen as in Theorem \ref{thm_conv}. Then, we have the
  following convergences in distribution.
  \begin{enumerate}   
  \item   
    For $k\in\N$,
    \begin{align*}
      \left( X^{(1)} - X^{(0)} ,\dots, X^{(k)} - X^{(0)} \big| \tilde{X}_n^{(0)} > - \log u(n)\right)
      \stackrel{d}\to \mathcal N(M,\Sigma), \quad n\to\infty,
    \end{align*}
    where $\mathcal N(M,\Sigma)$ denotes the multivariate normal
    distribution with mean vector $M =
    -\diag(\Psi_{k,(0,\dots,k)}(\Lambda))/2$ and covariance matrix $\Sigma =
    \Psi_{k,(0,\dots,k)}(\Lambda)$.
  \item
    For the bivariate case, i.e., $k=1$,
    \begin{align*}
      \left( X^{(1)} - X^{(0)} 
      \big| \tilde{X}_n^{(0)} > - \log u(n) \text{ or }  
      \tilde{X}_n^{(1)} > - \log u(n)\right)
      \stackrel{d}\to Z, \quad n\to\infty,
    \end{align*}
    where $Z$ is a real-valued random variable with density given by
    \begin{align*}
      g_\lambda(t) = \frac{1}{4\lambda\Phi(\lambda)}\phi\left(\lambda -
      \frac{|t|}{2\lambda}\right), 
      \quad t\in\R,\, \lambda = \lambda_{0,1}.
    \end{align*}
    Here, $\Phi$ and $\phi$ denote the standard normal distribution
    function and density, respectively.
  \end{enumerate}
\end{theo}

\begin{rem}
  \label{rem1}
  The positive definite matrix $\Sigma =\Psi_{k,(0,\dots,k)}(\Lambda)$ 
  contains all information on $\Lambda$. In fact, the transformation
  \[
  \Lambda(\Sigma) = \frac{1}{4}
  \left( \begin{array}{c|c}
    0 &  \diag(\Sigma)^\top\\[0.5em] \hline & \\[-.5em]
    \diag(\Sigma) & 
    \1 \diag(\Sigma)^\top + \diag(\Sigma)\1^\top -2\Sigma \\
  \end{array}\right)
  \]
  recovers the matrix $\Lambda = (\lambda^2_{i,j})_{0\leq i,j\leq k}$.
\end{rem}

\bigskip

Based on the convergence results in Theorem \ref{thm_HR}
we propose various estimation procedures for both
multivariate H\"usler-Reiss distributions (non-parametric case) 
and Brown-Resnick processes
with a parametrized family of variograms (parametric case).

\subsubsection{Non-parametric multivariate case}
 
For the likelihood based approach in \eqref{gen_likelihood} we first
consider the extremal set $A_1=(0, \infty)\times \R^k$ and put 
$N_1 = |I_{A_1}|$.
By part one of Theorem \ref{thm_HR} we have
\begin{align}
  -\log L(\Lambda; \sbf_1, \ldots, \sbf_n)
  &\approx -\log\left\{ \exp(-u(n))
    \frac{u(n)^{N_1}}{N_1!} \prod_{j=1}^{N_1} 
    \phi_{M(\Lambda), \Sigma(\Lambda)} \left( \sbf_{i_j} \right)\right\}\notag\\
  & \propto \frac{N_1}{2}\log\det\Sigma(\Lambda)
  + \frac{1}{2} \sum_{j=1}^{N_1} \left\{(\sbf_{i_j}-M(\Lambda))^\top
  \Sigma(\Lambda)^{-1}(\sbf_{i_j}-M(\Lambda))\right\},
  \label{loglike_PPP}
\end{align}
where $\sbf_{i}$ is the realization of $\Delta\Xbf_i$, $i=1, \ldots,
n$ and $\phi_{M(\Lambda), \Sigma(\Lambda)}$ is the density of the normal distribution with
mean vector $M(\Lambda)=-\diag(\Psi_{k,(0,\dots,k)}(\Lambda))/2$ and covariance matrix
$\Sigma(\Lambda)=\Psi_{k,(0,\dots,k)}(\Lambda)$.  The corresponding maximum likelihood
estimator is given by 
\begin{align}
  \hat\Lambda_{\MLEa} = \arg\min_{\Lambda\in\mathcal D}
  \left[ \frac{N_1}{2}\log\det\Sigma(\Lambda)+ \frac{1}{2} \sum_{j=1}^{N_1} 
  \left\{(\sbf_{i_j}-M(\Lambda))^\top\Sigma(\Lambda)^{-1}
  (\sbf_{i_j}-M(\Lambda))\right\} \right].\label{est_Lambda_MLE}
\end{align}
Notice that for this particular choice of $A$, the asymptotic value of
$\P( |I_{A_1}| = N_1 )$ does not depend on the parameter matrix $\Lambda$.
Hence, this ML ansatz coincides with simply maximizing the likelihood
of the increments without considering the number of points exceeding
the threshold. In the bivariate case, i.e., $k=1$ and $A_1=(0, \infty)\times\R$,
\eqref{loglike_PPP} simplifies to
\begin{align}
  -\log L(\lambda; s_1, \ldots, s_n)
  & \propto \frac{N_1\lambda^2}{2} + N_1\log \lambda + \frac{1}{8\lambda^2}\sum_{j=1}^{N_1} s_{i_j}^2,\label{like_MLE_biv}
\end{align} 
and the minimizer of \eqref{like_MLE_biv} can be given in explicit form:
\begin{align}
  \label{est_MLE_biv}
  \hat{\lambda}^2_{\MLEa} = \frac12\left\{-1 + \sqrt{ 1 + \frac{1}{N_1}\sum_{j=1}^{N_1} (\Delta \Xbf_{i_j})^2 } \right\}.
\end{align}

Staying in the bivariate case, for the choice $A_2 = [-\boldsymbol\infty,
  \0]^C$, we put $N_2 = |I_{A_2}|$ and by part two of Theorem \ref{thm_HR},
\begin{align*}
  -\log L(\lambda; s_1, \ldots, s_n)
  &\approx -\log\left\{\exp(-2\Phi(\lambda)u(n))
    \frac{(2\Phi(\lambda)u(n))^{N_2}}{N_2!} \prod_{j=1}^{N_2}
    g_\lambda(s_{i_j})\right\}\\
  & \propto 2\Phi(\lambda)u(n) +
  \frac{N_2\lambda^2}{2} +\frac{1}{8\lambda^2}\sum_{j=1}^{N_2} s_{i_j}^2.
\end{align*}
Numerical optimization can be applied to obtain the estimator
\begin{align*}
  \hat{\lambda}^2_{\MLEb} = \arg\min_{\theta\geq 0} 
  \left\{2\Phi(\sqrt{\theta})u(n) +
  \frac{N_2\theta}{2} +\frac{1}{8\theta}\sum_{j = 1}^{N_2} (\Delta\Xbf_{i_j})^2\right\}.
\end{align*}

\bigskip

While the above likelihood-based estimators (except for
\eqref{est_MLE_biv}) require numerical optimization, the following
approach is computationally much more efficient: A natural estimator
for $\Sigma =\Psi_{k,(0,\dots,k)}(\Lambda)\in\R^{k\times
  k}$ based on the first part of Theorem \ref{thm_HR} is given by the
empirical covariance $\hat{\Sigma}$ of the extremal increments
$\Delta\mathbf{X}_i =(X_{i}^{(1)} - X_{i}^{(0)},\dots, X_{i}^{(k)} -
X_{i}^{(0)})$ for $i\in I_{A_1}$, i.e.\
\begin{align}
  \label{sigma_ddim}
  \hat{\Sigma} = \frac{1}{N_1} \sum_{j = 1}^{N_1} (\Delta\mathbf{X}_{i_j} - \hat{\mu})(\Delta\mathbf{X}_{i_j} - \hat{\mu})^\top,
  \qquad \hat{\mu} = \frac{1}{N_1} \sum_{j = 1}^{N_1} \Delta\mathbf{X}_{i_j}.
\end{align}
By Remark \ref{rem1} this also gives an estimator
$\hat{\Lambda}_{\Var}=\Lambda(\hat\Sigma)$ for the parameter matrix
$\Lambda$, which we call the \emph{variance-based estimator}. 
Apart from its simple form,
another advantage of \eqref{sigma_ddim} is that
$\hat{\Sigma}$ is automatically a positive definite matrix and hence,
$\hat{\Lambda}_{\Var}$ is conditionally negative definite and therefore a valid
matrix for a $(k+1)$-variate H\"usler-Reiss distribution. Note
that \eqref{sigma_ddim} is not the maximum likelihood estimator (MLE)
for $\Sigma$ since the mean of the conditional distribution of
$\Delta\mathbf X_i$ depends on the diagonal of $\Sigma$.  The MLE of
$\Sigma$ is instead given by optimizing \eqref{est_Lambda_MLE}
w.r.t.\ $\Sigma$, which, to our knowledge, does not admit a closed
analytical form.  

Applying \eqref{sigma_ddim} with $k=1$ yields the bivariate 
variance-based estimator  
\begin{align}
  \label{bivvar}
  \hat{\lambda}^2_{\text{Var}} = \frac{1}{4N_1} 
  \sum_{j=1}^{N_1} ({X}_{i_j}^{(1)} - {X}_{i_j}^{(0)} - \hat{\mu})^2 ,
  \qquad \hat{\mu} = \frac{1}{N_1} \sum_{j=1}^{N_1} ({X}_{i_j}^{(1)} - {X}_{i_j}^{(0)}).
\end{align}
Since the mean of the extremal increments is also directly related to the parameter $\lambda$, another sensible estimator might be 
\begin{align}
  \label{bivmean}
  \hat{\lambda}^2_{\mean} 
  = -\frac{1}{2N_1} \sum_{j=1}^{N_1} ({X}_{i_j}^{(1)} - {X}_{i_j}^{(0)}).
\end{align}

\subsubsection{Parametric approach for Brown-Resnick processes}
\label{sec_increm_param}

Statistical inference for Brown-Resnick processes as in \eqref{BR_proc}
is usually based on fitting a parametric variogram model
$\{\gamma_\vartheta : \vartheta\in\Theta\}$, $\Theta\subset\R^j$, $j\in\N$, to
point estimates of the extremal coefficient function \eqref{ECF} based on
the madogram. Alternatively, composite likelihood approaches are
used in connection with block maxima of bivariate data (\citet{dav2012}).\\
Since for $t_0,\dots, t_k\in\R^d$, the vector $(\xi(t_0), \ldots, \xi(t_k))$ with $\xi$ being a
Brown-Resnick process associated to the variogram $\gamma:\R^d \to [0,
  \infty)$ is H\"usler-Reiss distributed with parameter matrix
\begin{align*}
  \Lambda=(\gamma(t_i-t_j)/4)_{0\leq i, j\leq k},
\end{align*}
the above estimators enable parametric estimation of Brown-Resnick
processes. In fact, replacing $\Lambda$ in \eqref{est_Lambda_MLE} by 
\begin{align}
  \label{lambda_theta}
  \Lambda(\vartheta) =
  \left(\gamma_\vartheta(t_i-t_j)/4\right)_{0\leq i, j\leq k}
\end{align}
leads to the ML estimator 
\begin{align*}
\hat\vartheta_{\MLEa} = \arg\min_{\vartheta\in\Theta}\left\{
-\log L(\Lambda(\vartheta); \sbf_1, \ldots, \sbf_n)\right\}
\end{align*}
with $L$ as in \eqref{loglike_PPP}. Note that, other than in classical
extreme value statistics, here the use of higher dimensional densities 
is feasible and promises a gain in accuracy.\\ 
Estimation of $\vartheta$ can also be based on any of the bivariate
estimators $\hat\lambda^2_{\MLEa}$,
$\hat\lambda^2_{\MLEb}$, $\hat\lambda^2_{\Var}$, $\hat\lambda^2_{\mean}$, 
or on the
multivariate estimator $\hat\Lambda_{\Var}$ by ``projecting'' the latter
matrix or the matrix consisting of all bivariate estimates onto the
set of matrices $\bigl\{ (\gamma_\vartheta(t_i-t_j)/4)_{0\leq i, j\leq k}
$:$\vartheta\in\Theta\bigr\}$, i.e.,
\begin{align}
  \label{est_proj}
  \hat\vartheta_{\text{PROJ}} = \arg\min_{\vartheta\in\Theta} \left\| \left(\hat\lambda_{ij}^2-\gamma_\vartheta(t_i-t_j)/4\right)_{0\leq i, j\leq k} \right\| ,
\end{align}
where $\|\cdot\|$ can be any matrix norm.\\
Similar to \citet{bac2012}, the bivariate estimators can readily be used 
in a parametric composite likelihood framework.

\subsection{Inference based on spectral densities}
\label{spec_inference}

As at the beginning of Section \ref{sec_estimation}, let
$\mathbf{X}_{i}$, $i = 1,\dots, n$, be a sequence of independent
copies of $\Xbf$, already standardized to exponential margins, in the
MDA of the max-stable distribution $H_\Lambda$. Since we work in the
spectral domain in this section, we will switch to standard Fr\'echet
margins with distribution function $\exp(-1/y), y\geq 0$.  More
precisely, we consider the vectors $\Ybf=\exp (\Xbf)$ and $\Ybf_{i} =
\exp (\Xbf_{i})$, $i = 1,\dots, n$, which are in the MDA of the
H\"usler-Reiss distribution $G_\Lambda(\mathbf{x}) =
H_\Lambda(\log\mathbf{x}), \mathbf{x} \geq \0$, with standard
Fr\'echet margins.\\
The most convenient tool to
characterize the dependence structure of a multivariate extreme value
distribution  is via its spectral measure. To this end, let $\tilde{\Ybf}_{n} = \Ybf / n$
and $\tilde{\Ybf}_{i,n} = \Ybf_i / n$
denote the rescaled data such that the point process $P_n = \sum_{i=1}^n \delta_{\tilde{\Ybf}_{i,n}}$
converges, as $n\to \infty$, to a non-homogeneous Poisson point
process $P$ on $[0,\infty)^{k+1}\setminus \{\0\}$ with intensity measure
  $\nu([\mathbf{0},\mathbf{x}]^C) = -\log G_\Lambda(\mathbf{x})$. Transforming a
  vector $\mathbf{x}=(x_0,\dots,x_k)\in [0,\infty)^{k+1}\setminus \{\0\}$
    to its pseudo-polar coordinates
\begin{align}
  \label{polar}
  r = \|\mathbf{x}\|, \qquad \boldsymbol{\omega} = r^{-1} \mathbf{x},
\end{align}
for any norm $\| \cdot \|$ on $\R^{k+1}$, we can rewrite $\nu$ as a
measure on $(0,\infty)\times S_{k}$, where $S_{k}$ is the
$k$-dimensional unit simplex $S_{k} = \{ \mathbf{y} \geq \0: \|y\| = 1
\}.$ Namely, we have $\nu(d\mathbf{x}) = r^{-2}dr \times
M(d\boldsymbol{\omega})$, where the measure $M$ is called the spectral
measure of $G_\Lambda$ and embodies the dependence structure of the
extremes. For our purposes, it is most convenient to choose the
$L_1$-norm, i.e., $\| \mathbf{x} \|_1 = \sum_{i=0}^k |x_i|$. In this
case, for the set $A_{r_0} = \{ \mathbf{x}\in
[0,\infty)^{k+1}\setminus \{\0\}: \| \mathbf{x} \|_1 > r_0 \}$,
  $r_0>0$, we obtain
\begin{align}
  \label{Ar0}
  \nu(A_{r_0}) = r_0^{-1} M(S_{k}) = r_0^{-1} \cdot (k+1),
\end{align}
since the measure $M$ satisfies $\int_{S_{k}} \omega_i \ M(d\boldsymbol{\omega}) = 1$
for $i=0,\dots, k$.
Hence, the $\nu$-measure of $A_{r_0}$ does not depend on the
parameters of the specific model chosen for $M$.
The distribution function can be written as 
\begin{align*}
  G_\Lambda(\mathbf{x}) = \exp\left\{ -\int_{S_{k}} \max\left(\frac{\omega_0}{x_0},\dots,\frac{\omega_k}{x_k}\right) M(d\boldsymbol{\omega})\right\},\quad
  \mathbf{x} \geq \0.
\end{align*}

As the space of all spectral measures is infinite dimensional, there
is a need of parametric models which are analytically tractable and at
the same time flexible enough to approximate the dependence
structure in real data sufficiently well. 
Parametric models are usually given in terms of their spectral
density $h$ of the measure $M$. The book by \citet{kot2000} gives an
overview of parametric multivariate extreme value distributions, most
of them, however, being only valid in the bivariate case. For the
multivariate case only few models are known, e.g., the logistic
distribution and its extensions \citep{taw1990,joe1990} and the
Dirichlet distribution \citet{col1991}. The recent interest in this topic 
resulted in new multivariate
parametric models (\citet{bol2007,coo2010}) as well as in general
construction principles for multivariate spectral measures
(\citet{bal2011}). All these approaches have in common that they
propose models for multivariate max-stable distributions in order to
fit data obtained by exceedances over a certain threshold or by
block maxima. \\
Given a parametric model for the spectral density
$h(\,\cdot\,;\vartheta)$, we have the analog result as in Theorem
\ref{thm_conv} for the Fr\'echet case with $A = A_{r_0}$ and
$g:\R^{k+1} \to S_{k}, \mathbf{x} \mapsto \mathbf{x}/
\|\mathbf{x}\|_1$, which now satisfies the multiplicative invariance
property $g(a\cdot \mathbf{x}) = g(\mathbf{x})$, for all $a \in\R$.
The Fr\'echet version of \eqref{conv_assump} for this choice of $g$ and $A$
reads as
\begin{align}  
  \lim_{n\to\infty}\P\left\{\Ybf / \|\Ybf\|_1 \in B\, |\, \tilde{\Ybf}_n \in A_{r_0} /  u(n) \right\} 
  = \frac{M(B)}{M(S_k)} = \frac{1}{k+1}\int_{B} h(\boldsymbol{\omega};\vartheta) d\boldsymbol{\omega}, 
\end{align}
for all $B \in \mathcal{B}(S_k)$ and $u(n)$, $n\in\N$, as in Theorem
\ref{thm_conv}. Based on this conditional distribution of those
$\Ybf_{i}$ for which the sum $\|\Ybf_{i}\|_1$ is large, similarly to
\eqref{gen_likelihood} we obtain the likelihood
\begin{align}   
  \notag L_{A_{r_0}}\bigl(\vartheta;& (r_1, \boldsymbol \omega_1), \ldots, (r_n, \boldsymbol \omega_n)\bigr) \\
&\approx 
  \exp\{-\nu(A_{r_0}/u(n))\}\frac{\nu(A_{r_0}/u(n))^{|I_0|}}{|I_0|!}\prod_{i\in I_0}
  r_i^{-2}(k+1)^{-1}h(\boldsymbol{\omega}_i;\vartheta)\notag\\
  \label{spec_likelihood} &\propto \prod_{i\in I_0}
  h(\boldsymbol{\omega}_i;\vartheta),
\end{align}
where  $\{
(r_i,\boldsymbol{\omega}_i) : 1\leq i\leq n \}$ are the pseudo-polar
coordinates of $\{\tilde{\Ybf}_{i,n} : 1\leq i\leq n\}$ as in
\eqref{polar} and $I_0$ is the set of all indices $1\leq i\leq n$ with
$\tilde{\Ybf}_{i,n}\in A_{r_0}/u(n)$. Note that the proportional part in
\eqref{spec_likelihood} only holds because the $\nu$-measure of $
A_{r_0}$ is independent of the model parameter~$\vartheta$.\\ For the
H\"usler-Reiss distribution it is possible to write down 
the spectral density $h(\, \cdot \,;\Lambda)$ explicitly.
\begin{prop}
  \label{prop_spec}
  For any matrix $\Lambda = \left( \lambda^2_{i,j}\right)_{0\leq
   i,j \leq k} \in\mathcal{D}$ the H\"usler-Reiss distribution can be written as 
  \begin{align*}
    G(\mathbf{x}) = \exp\left\{ -\int_{S_{k}} \max\left(\frac{\omega_0}{x_0},\dots,\frac{\omega_k}{x_k}\right) h(\boldsymbol{\omega};\Lambda) \,d\boldsymbol{\omega} \right\},
  \end{align*}
  with spectral density
  \begin{align}  
    \label{spec_den}
    h\left(\boldsymbol{\omega}, \Lambda \right) = \frac{1}{\omega_0^2\cdots
      \omega_k (2\pi)^{k/2}
      |\det\Sigma|^{1/2}} \exp\left( -\frac12
    \tilde{\boldsymbol{\omega}}^\top \Sigma^{-1}\tilde{\boldsymbol{\omega}}\right),\quad \boldsymbol{\omega}\in S_k,
\end{align}
  where $\Sigma = \Psi_{k,(0,\dots,k)}(\Lambda)$ and $\tilde{\boldsymbol{\omega}} = (\log \frac{\omega_i}{\omega_0} + 2\lambda^2_{i,0}: 1\leq i \leq k)^\top$.
\end{prop}

\subsubsection{Non-parametric, multivariate case}

Based on the explicit expression for the spectral density of the 
H\"usler-Reiss distribution in~\eqref{spec_den}, we define the estimator 
$\hat{\Lambda}_{\SPEC}$ of $\Lambda$ as the matrix in $\mathcal{D}$ that
maximizes the likelihood in \eqref{spec_likelihood}, i.e.,
\begin{align}
  \label{est_spec}
  \hat{\Lambda}_{\SPEC} = \arg\min_{\Lambda \in\mathcal{D}} \left(\frac{|I_0|}{2} \log\det\Psi_{k,(0,\dots,k)}(\Lambda) + 
  \frac12 \sum_{i\in I_0}
  \tilde{\boldsymbol{\omega}_i}^\top \Psi_{k,(0,\dots,k)}(\Lambda)^{-1}\tilde{\boldsymbol{\omega}_i} \right).
\end{align}
In the bivariate case, the spectral density in \eqref{spec_den} 
simplifies to 
\begin{align*}
h(\boldsymbol \omega; \lambda) = 
\frac{1}{ 2\lambda\omega_0^2 \omega_1 (2\pi)^{1/2}} \exp\left(-\frac{(\log \frac{\omega_1}{\omega_0}+2\lambda^2)^2}{8\lambda^2}\right) 
\end{align*} 
and the corresponding estimator can be given in explicit form:
\begin{align}
  \label{est_spec_biv}
  \hat{\lambda^2}_{\SPEC} 
  = \frac12\left[-1 + 
  \sqrt{ 1 + \frac{1}{|I_0|}\sum_{i\in I_0} \left\{\log \left(\tilde\Ybf_i^{(1)}\big/ \tilde\Ybf_i^{(0)}\right)\right\}^2} \right].
\end{align}
Note that the estimators \eqref{est_spec} and \eqref{est_spec_biv} have 
exactly the same form
as the maximum likelihood estimators \eqref{est_Lambda_MLE} and 
\eqref{est_MLE_biv}, respectively, for the
extremal increments. However, the specification of the set $A$ differs and
so does the choice of extreme data that is plugged in.

\subsubsection{Parametric approach for Brown-Resnick processes}
Analogously to Section \ref{sec_increm_param}, we obtain a parametric
estimate of the dependence structure of a Brown-Resnick
process based on a parametric family of variograms by replacing
$\Lambda$ on the right-hand side of \eqref{est_spec} by
$\Lambda(\vartheta)$ defined in \eqref{lambda_theta}.
This yields 
\begin{align}
\hat\vartheta_{\SPEC} = \arg\min_{\vartheta\in\Theta}\left\{
-\log L_{A_{r_0}}(\Lambda(\vartheta); (r_1, \boldsymbol \omega_1), \ldots, (r_n, \boldsymbol \omega_n))\right\}\label{est_spec_param}.
\end{align}

\iffalse
\subsection{Density of bivariate H\"usler-Reiss distribution 
(section to be deleted?)}
The density is given by
\begin{align*}
  \frac{\partial^2}{\partial x \partial y}& \exp( - \mu( (-\infty,
  (x,y)]^C ) )\\ &= \exp( - \mu( (-\infty, (x,y)]^C )
  )e^{-x}\left[\Phi\left(\lambda + \frac{y-x}{2\lambda}
    \right)\Phi\left(\lambda + \frac{x-y}{2\lambda} \right)e^{-y} +
    \phi\left(\lambda + \frac{y-x}{2\lambda} \right)
    \frac{1}{2\lambda}\right].
\end{align*} 
\fi

\section{Simulation study}\label{sim_study}

We compare the performance of the 
different parametric and non-parametric estimation 
procedures of Brown-Resnick processes and H\"usler-Reiss distributions 
proposed in the previous section via a simulation study.

\begin{figure}[!htb]
\caption{\label{fig:results}Estimated extremal coefficients compared to the true ones of bivariate H\"usler-Reiss distributions. 500 repetitions. Block size for block maxima is 150. Left: $\hat\theta$ vs.\ $\lambda^2$. Right: relative difference of $\hat\theta$ to the true value of $\theta$.}
\begin{center}
  \includegraphics[width=1\textwidth, page=1, clip=true]{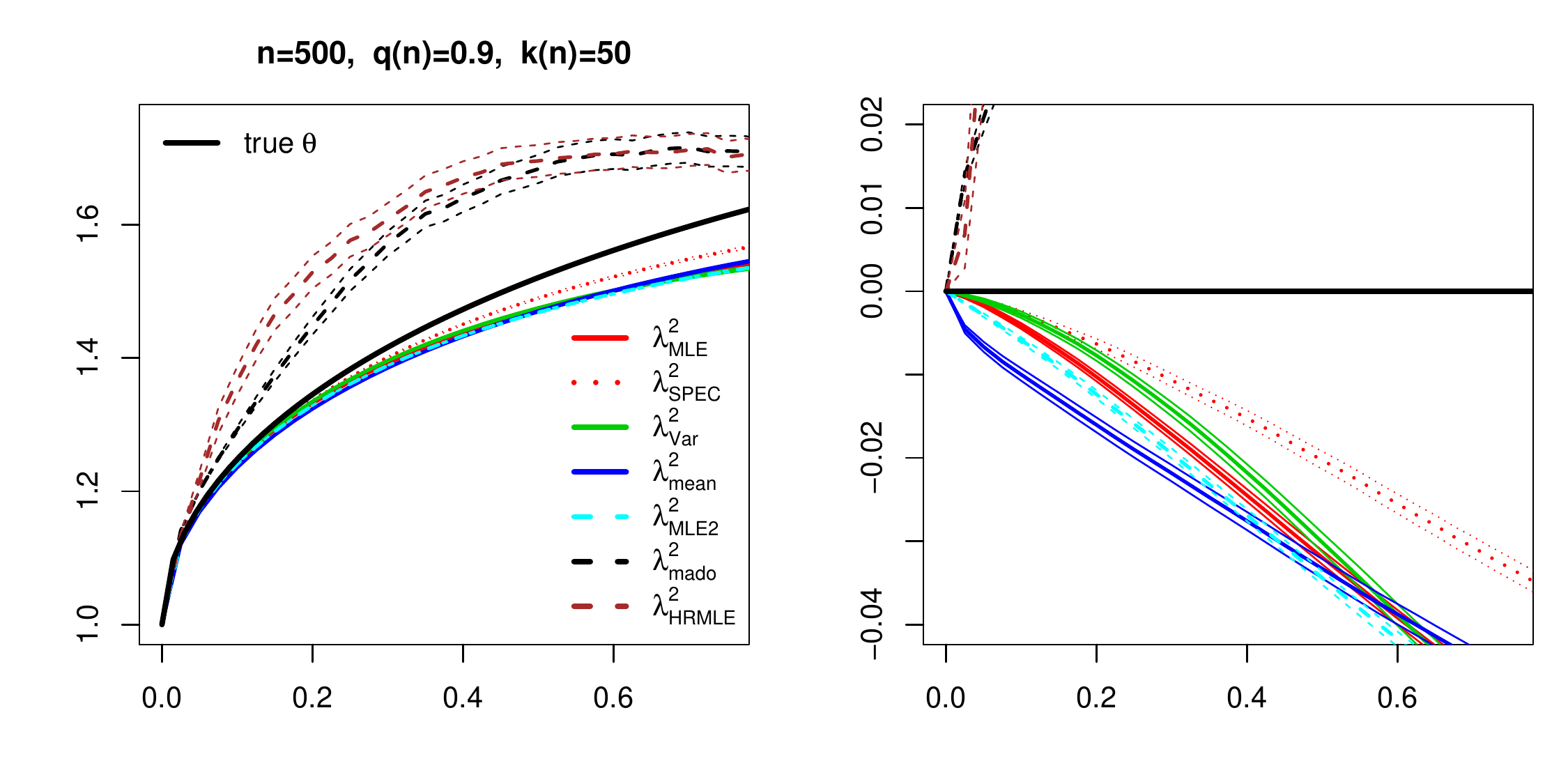}\\[-1.5em]
  \includegraphics[width=1\textwidth, page=2, clip=true]{consistency.proof13.alpha1}\\[-1.5em]
 \includegraphics[width=1\textwidth, page=3, clip=true]{consistency.proof13.alpha1}
\end{center}
\end{figure}

In the first instance, we consider bivariate data that is in the MDA
of the H\"usler-Reiss distribution with known dependence parameter
$\lambda=\lambda_{0,1}$. For simplicity, we simulate data from the
H\"usler-Reiss distribution itself, which does not mean that the
thresholding procedure via the set $A$ becomes obsolete. All
estimators rely on considering only extremal events and hence, there
is no obvious advantage over using any other data being in the MDA of
$H_\lambda$. We compare the estimators $\hat\lambda^2_{\MLEa}$,
$\hat\lambda^2_{\MLEb}$, $\hat\lambda^2_{\Var}$,
$\hat\lambda^2_{\mean}$ and $\hat\lambda^2_{\SPEC}$ from
Section~\ref{sec_estimation} for different sample sizes $n\in\{500,
8000, 100000\}$.  The sequence of thresholds $u(n)$ is chosen in such
a way that the number of exceedances $k(n)$ increases to $\infty$, but
at the same time, the corresponding quantile $q(n) = 1-k(n)/n$
approaches $1$, as $n\to \infty$.  In addition to the new threshold
based estimators, we include the classical estimators, which use block
maxima, namely the madogram estimator $\hat\lambda_{\mado} =
\Phi^{-1}(\hat\theta_{\mado}/2)$ \citep{coo2006} and the ML
estimator $\hat\lambda^2_{\HRMLE}$ of the bivariate
H\"usler-Reiss distribution. To model a year of (dependent) data, we
we choose a block size of 150 which is of order of but less than
365.\\ The
pseudo-code of the exact simulation setup is the following: \texttt{
\begin{enumerate}
\item for $\lambda^2\in\{k\cdot 0.025 : k=1, \ldots, 30\}$ \label{simustep1}
\item \qquad for $n\in\{500, 8000, 100000\}$ 
  \label{simustep2}
\item \qquad\qquad simulate $n$ bivariate H\"usler-Reiss distributions \\[-2em]
\item[] \qquad\qquad with parameter
  $\lambda$
\item \qquad\qquad for $\hat\lambda^2\in\Bigl\{\hat{\lambda}^2_{\MLEa},
  \hat{\lambda}^2_{\MLEb},\hat\lambda^2_{\Var},
  \hat\lambda^2_{\mean}, \hat\lambda^2_{\SPEC}, 
  \hat\lambda^2_{\mado}, \hat\lambda^2_{\HRMLE}\Bigr\}$
\item \qquad\qquad\qquad estimate $\lambda^2$ through
  $\hat\lambda^2$
\item \qquad\qquad\qquad obtain an estimate of the corresponding
  extremal \\[-2em]\label{simustep5}
\item[]\qquad\qquad\qquad  coefficient $\theta$ 
through $\hat\theta =
  \theta(\hat\lambda) = 2\Phi(\hat\lambda)$
\item repeat  (\ref{simustep1})-(\ref{simustep5})
  500 times \label{simustep6}
\end{enumerate}} 
\noindent Since the finite dimensional margins of a Brown-Resnick process are
H\"usler-Reiss distributed, we can easily implement step
(\ref{simustep1}) by simulating a one-dimensional Brown-Resnick
process with variogram $\gamma(h)=|h|$ on the interval $[0, 3]$.
Since we consider bivariate H\"usler-Reiss distributions for different
values of $\lambda^2$ lying on a fine grid, we visualize the estimates
$\hat{\theta}$ as functions of the true $\lambda^2$
(Figure~\ref{fig:results}). However, it is important to remark that
estimation in this first part of the study is exclusively based on the
bivariate distributions. For each value of $\lambda^2$, we repeat
simulation and estimation 500 times. Figure \ref{fig:results} shows
the pointwise mean value of the extremal coefficient and the
corresponding empirical $95\%$ confidence intervals. \\ As expected,
in finite samples, all estimators based on multivariate POT methods
underestimate the true degree of extremal dependence since they are
based on an asymptotic distribution with non-zero mean while the
simulated data come from a stationary process.  As the sample size $n$
and the threshold $u(n)$ increase, all estimators approach the true
value. Among the POT-based estimators, $\hat\lambda^2_{\SPEC}$ seems
to be at least as good as the other estimators, uniformly for all
values of $\lambda^2$ under consideration.  $\hat\lambda^2_{\Var}$
performs well for small values of $\lambda^2$ but is more biased than
other estimators for large values of $\lambda^2$. The good performance
of $\hat\lambda^2_{\mean}$ for large values of $\lambda^2$ might
be due to the fact that it only uses first moments of the extremal
increments and is hence less sensible to aberration of the finite
sample distribution from the asymptotic distribution. Compared to the
two estimators based on block maxima, the POT-based estimators all
perform well even for small data sets, which is a great advantage for
many applications. Moreover, the variances of the POT-estimates are
generally smaller than those based on block maxima, since more data
can be used. Finally, note that the POT-based estimation does not
exploit the fact, that the simulated data in the max-domain of attraction
 is in fact the
max-stable distribution itself. The speed of convergence may though
differ when using data from other models in the MDA. In contrast,
$\hat\lambda^2_{\mado}$ and $\hat\lambda^2_{\HRMLE}$
do profit from simulating i.i.d.\ realizations of the 
max-stable distribution itself since then,
the blockwise maxima are exactly H\"usler-Reiss distributed and not
only an approximation as in the case of real data.
 
\begin{figure}[!htb]
\caption{\label{fig:results_param_aa}Parametric fit of Brown-Resnick process.}
\begin{center}
  \includegraphics[width=1\textwidth, page=1, clip=true]{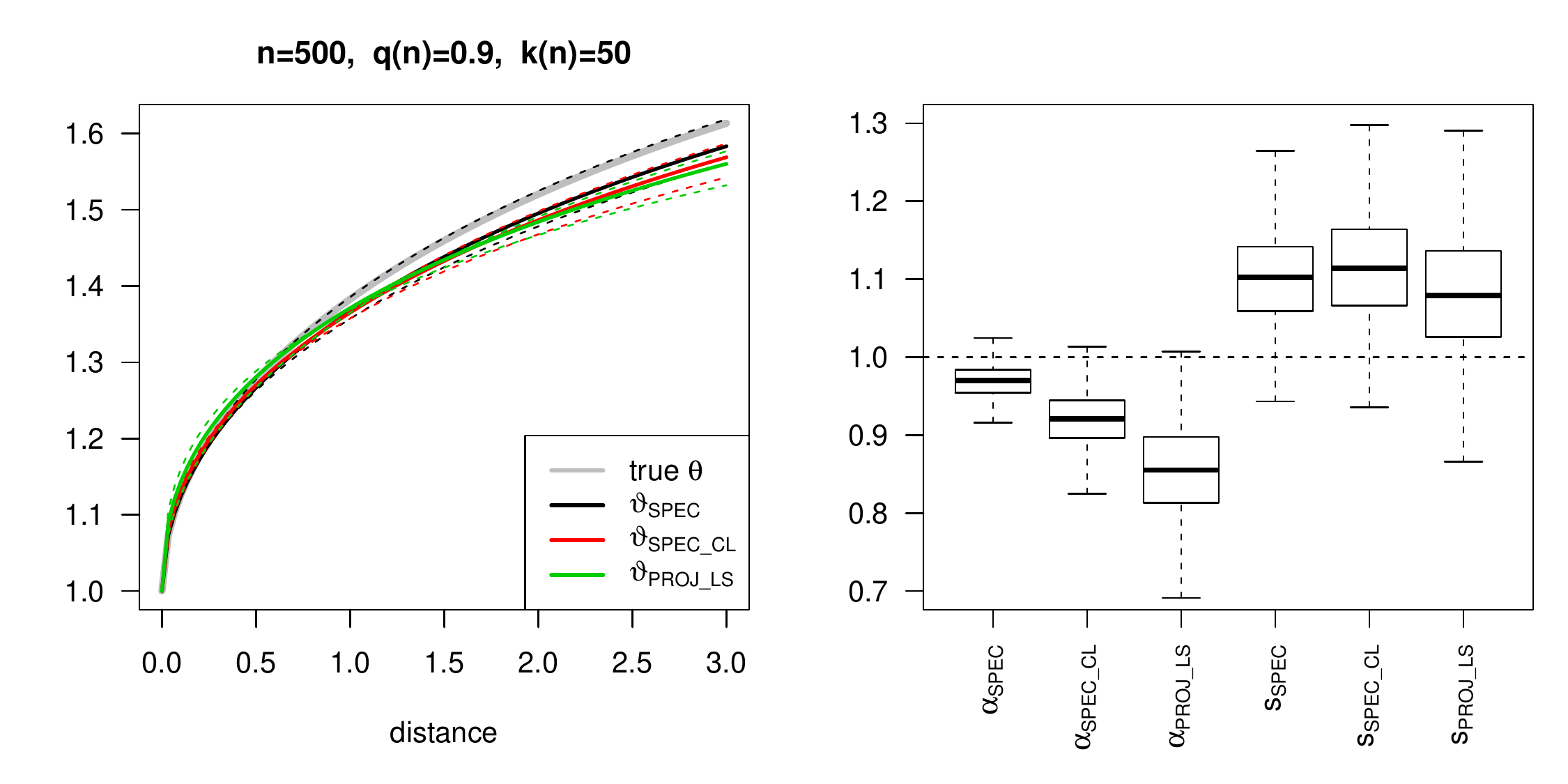}
  \includegraphics[width=1\textwidth, page=2, clip=true]{consistency.proof13parametric.alpha1}
 \includegraphics[width=1\textwidth, page=3, clip=true]{consistency.proof13parametric.alpha1}
\end{center}
\end{figure}

In the second part of the simulation study we examine the performance
of parametric estimates of Brown-Resnick processes using the same data
as above. While the true variogram is $\gamma(h)=|h|$, we estimate the
 parameter vector $(\alpha, s)$ for the family of
variograms $\gamma_{\alpha, s}(h) = \|h/s\|^\alpha$, $\alpha\in (0,
2]$, $s>0$. We compare the following three estimators: the spectral
estimator $\widehat{(\alpha, s)}_{\SPEC}$, given by
\eqref{est_spec_param} and using the full multivariate density; the
composite likelihood estimator $\widehat{(\alpha, s)}_{\SPEC,
  \text{ CL}}$, defined as the maximizer of the product of all
bivariate spectral densities, implicitly assuming independence of all
tuples of locations; and the least squares estimator $\widehat{(\alpha,
  s)}_{\PROJ, \text{ LS}}$, given by \eqref{est_proj} for the Euclidean
norm, where $\hat\lambda^2_{\MLEa}$ serves as non-parametric input. The
estimated values of $\alpha$ and $s$ are compared in the right column
of Figure \ref{fig:results_param_aa}. The left panel shows the
corresponding extremal coefficient functions for $\alpha$ and $s$
representing the mean, the $5\%$ sample quantile and the $95\%$ sample
quantile from the 500 repetitions, respectively.\\ The estimator
$\widehat{(\alpha, s)}_{\SPEC}$, which incorporates the full
multivariate information, performs best both in the sense of minimal
bias and minimal variance. Especially estimation of the shape
parameter of the variogram gains stability when using
higher-dimensional densities.  The projection estimator seems to have
the largest bias and the largest variance.  The results remain very
similar if we replace $\hat\lambda^2_{\MLEa}$ by one of the other
non-parametric estimators. Let us finally remark that all three
estimators can be modified by considering only small distances for
inference. Then, since the approximation error of the asymptotic
conditional distribution decays for smaller distances, this can
substantially improve the accuracy in a simulation framework, but
might distort the results in real data situations.

\clearpage

\section{Application: Wind speed data}\label{wind_data}

\begin{figure}[!Htb]
  \caption{\label{fig:locations}Left: locations of the 35
    meteorological stations.  Right:
    locations of the 25 non-coast stations before and 
    after multiplication with
    the anisotropy matrix $V(\hat\beta, \hat c)$. }
\begin{center}
  \includegraphics[width=0.49\textwidth]{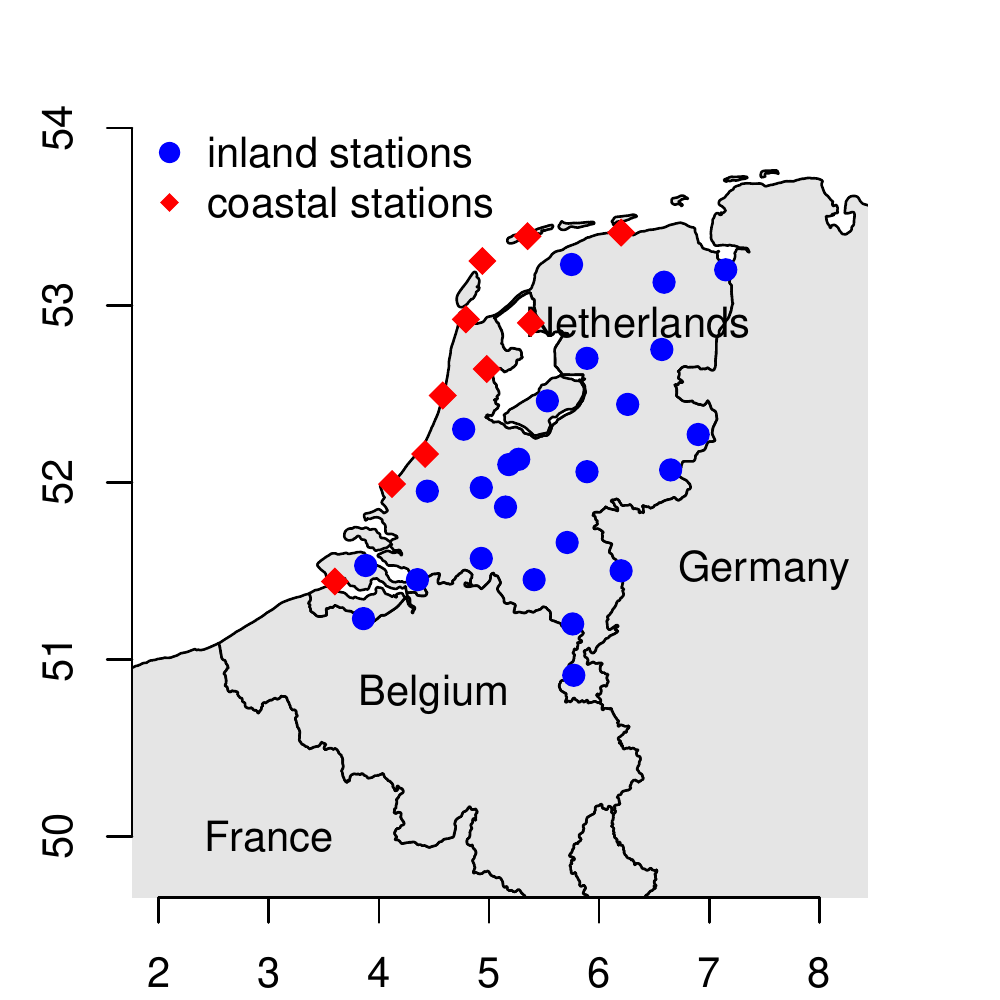}
  \includegraphics[width=0.49\textwidth]{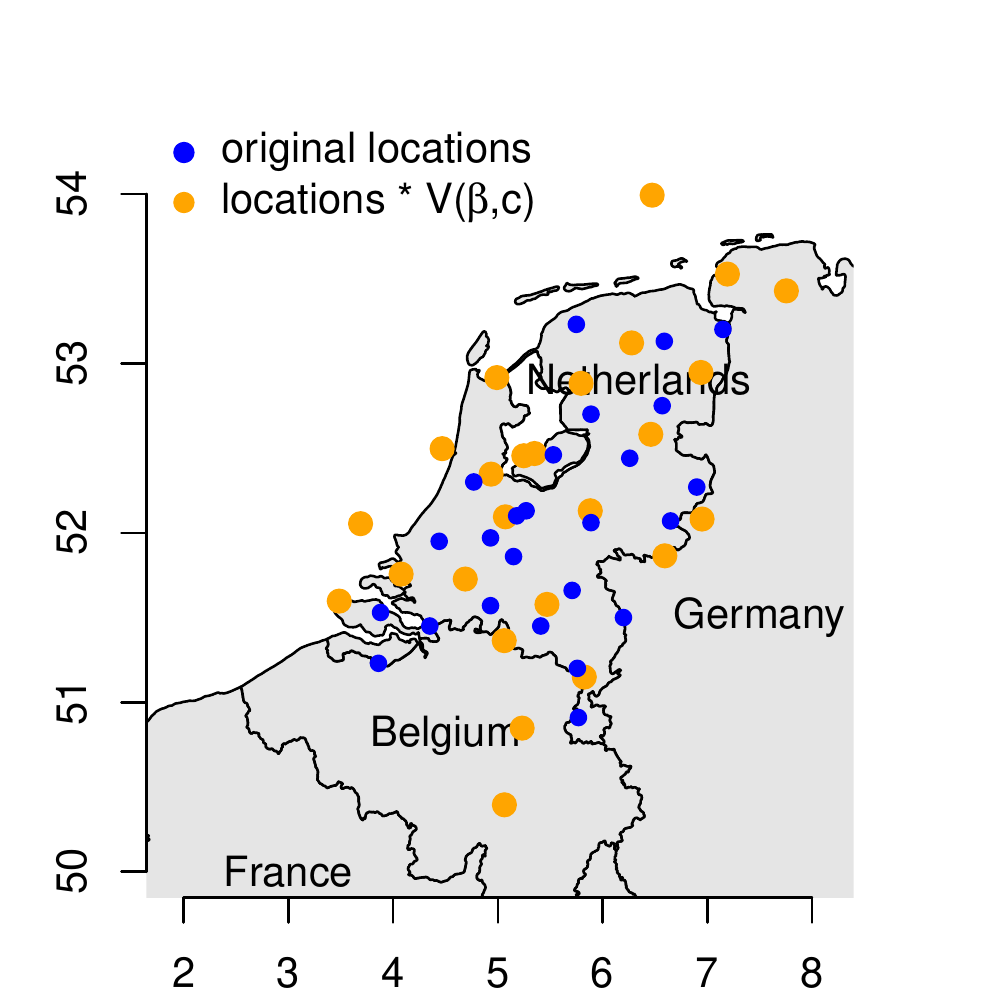}
\end{center}
\end{figure}

We apply the above theory of estimating H\"usler-Reiss distributions
to wind speed data provided by the Royal Netherlands Meteorological
Institute.  We use the daily maxima of wind speeds measured at 35
meteorological stations $x_1, \ldots, x_{35}\in\Xcal$, where the
set $\Xcal\subset \R^2$ denotes the geographical coordinates of the Netherlands.
The data cover a 23-year period of 8172 days
from 1990/01/01 to 2012/05/12. Figure \ref{fig:locations} provides an
overview of the spatial locations of the stations. 

\subsection{Stationarity assumption with zonal anisotropy}
\label{wind_stationary}

\begin{figure}[!htb]
\caption{\label{fig:fitBR.coast.aniso}Estimated extremal coefficients based on
  $\hat\lambda^2_{\MLEa}$ against distance between the stations. Left
  panel: original locations with and without coast stations.  
  Middle panel: transformed locations
  (only non-coast stations), extremal coefficient functions
  corresponding to the parameters in  Table \ref{tab:results} are
  included. Right panel: comparison to madogram estimator.  
}
\begin{center}
\includegraphics[width=
    0.33\textwidth]{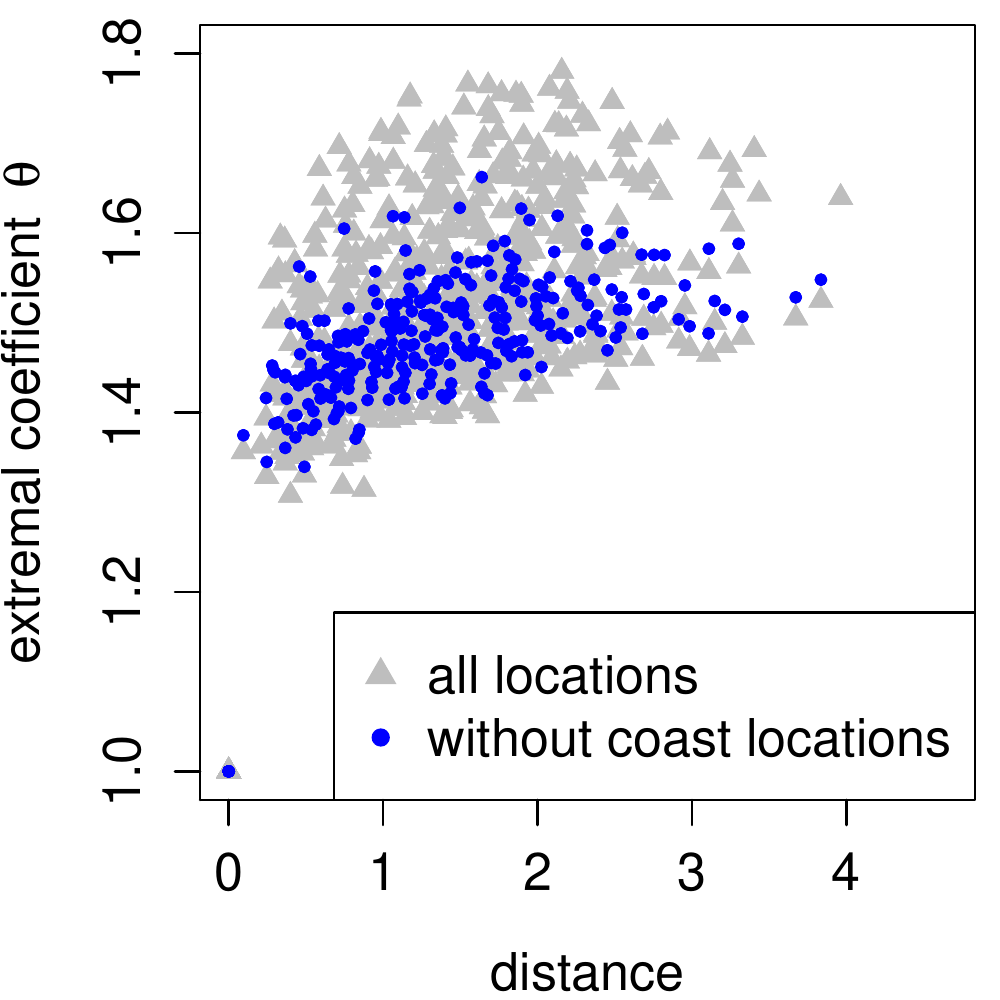}%
\includegraphics[width=
    0.33\textwidth]{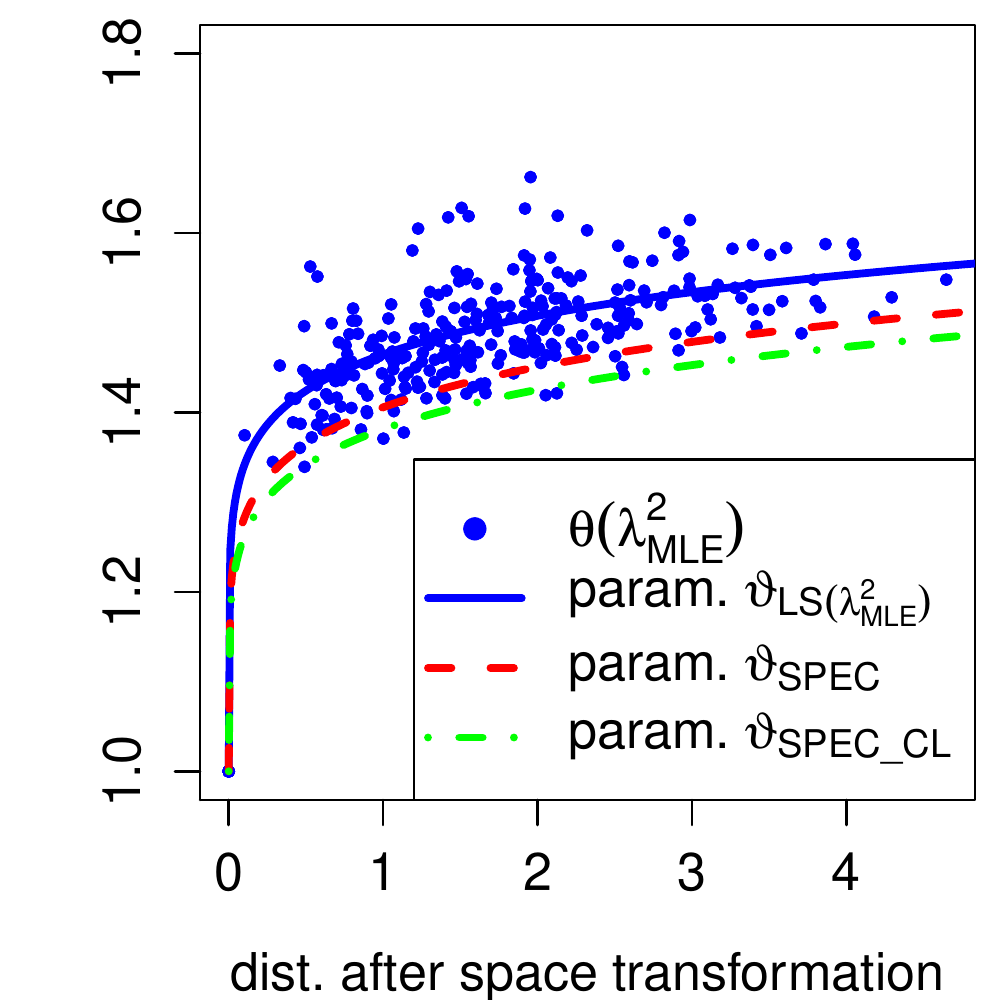}%
\includegraphics[width=
    0.33\textwidth]{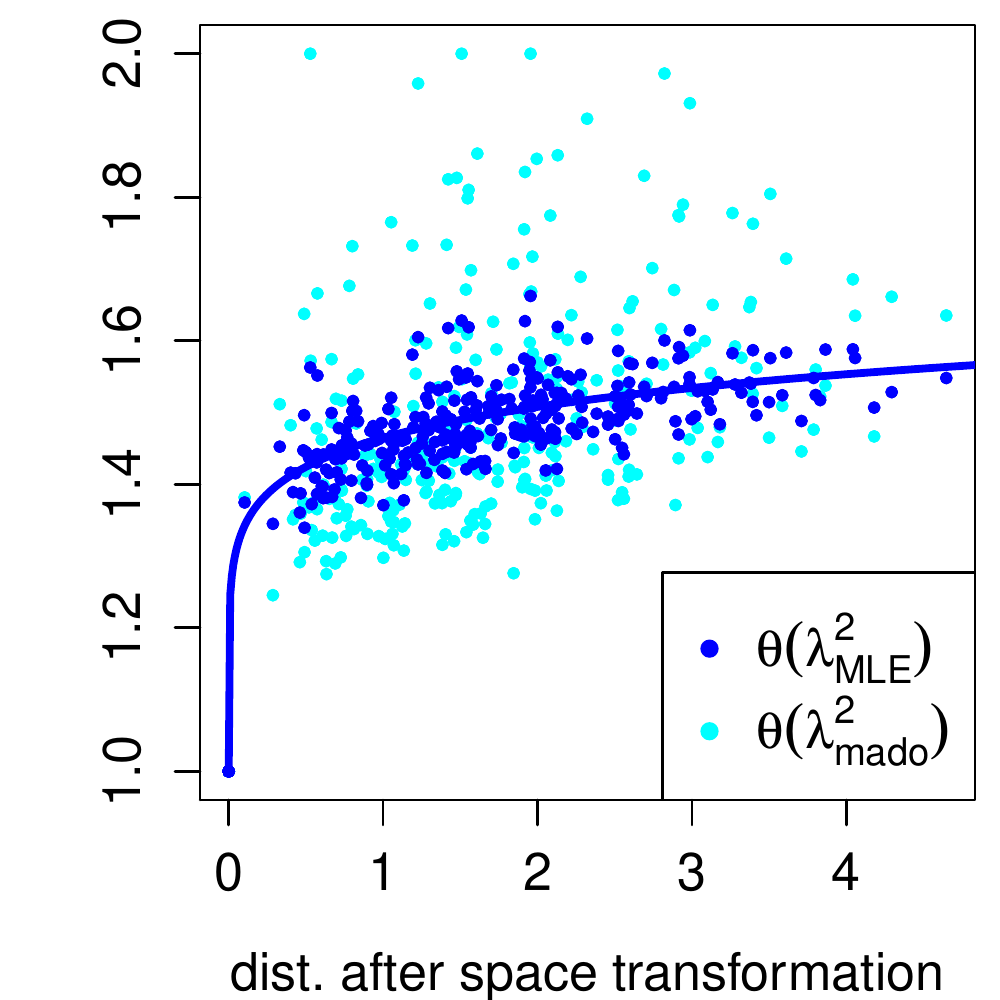}% 
\end{center}
\end{figure}

In the sequel, we use the data to fit a stationary Brown-Resnick
process based on the parametric family of variograms $\gamma_{\alpha, s}(h) =
\|h/s\|^\alpha$, $\alpha\in (0, 2]$, $s>0$. As mentioned in Section \ref{sec_BR},
this subclass of Brown-Resnick processes is a natural choice, since 
they arise as the max-limits of suitably rescaled, stationary Gaussian 
random fields. The stationarity assumption, however, turns out to
be unrealistic, since stations close to the coast exhibit
weaker extremal dependence to neighboring stations than inland stations.
This is illustrated in the left panel of Figure \ref{fig:fitBR.coast.aniso},
where the estimated bivariate extremal coefficients based on $\hat{\lambda}^2_{\MLEa}$ of all stations are compared
to those without the coastal stations. Hence, we restrict our analysis
to the 25 inland stations, say $T=\{x_1, \ldots, x_{25}\}$, when fitting a stationary
Brown-Resnick process. We therefore need to estimate
the shape parameter $\alpha$ and the scale parameter $s$ of the
corresponding parameter matrix $\Lambda_{\alpha, s}$ of the Brown-Resnick process
on $T$, given by $\Lambda_{\alpha, s}=\bigl(\gamma_{\alpha, s}(x_i-x_j)/4\bigr)_{1\leq i, j \leq 25}$.\\  
While the above class of variograms assumes isotropy of the underlying
process, meteorological data and particularly wind speed data can be
expected to exhibit a main direction of spatial dependence. We capture
this anisotropy by introducing a transformed space $\tilde\Xcal =
V\Xcal$ (cf.\ right panel of Figure \ref{fig:locations}), where
\begin{align*}
V=V(\beta, c) =
\left(\begin{smallmatrix} \cos\beta & -\sin\beta \\ c \sin\beta &
  c\cos\beta \end{smallmatrix}\right), \qquad \beta\in [0,2\pi],\ c > 0,
\end{align*} 
is a rotation and dilution matrix; \citet{bla2011} recently applied this 
idea to the extremal Gaussian process of \citet{sch2002}.
The new parametric variogram model becomes $\Lambda_{\vartheta}=
\bigl(\gamma_{\alpha, s}(Vx_i-Vx_j)/4\bigr)_{1\leq i, j \leq 25}$,
where $\vartheta=(\alpha, s, \beta, c)$ is the vector of parameters.
As in the above simulation study, we apply the three estimators
\begin{align}
\hat\vartheta_{\PROJ, \text{ LS}} = 
\arg\min_{\vartheta\in\Theta} \big\| (\hat\lambda^2_{\MLEa, ij})_{1\leq i, j\leq 25} - 
\Lambda_{\vartheta} \big\|_2,\qquad
\hat\vartheta_{\SPEC},\qquad
\hat\vartheta_{\SPEC, \text{ CL}}. \label{three_est}
\end{align} 
For all estimators, the data is first normalized as described at the
beginning of Section \ref{sec_estimation} and the threshold $u(n)$ is
chosen in such a way that, out of the 8172 days, all data above the
$97.5\%$-quantile are labeled as extremal. Note that these numbers
coincide with the second set of parameters $(n, q(n))$ in the
simulation study. Hence, the middle row of Figure~\ref{fig:results}
provides a rough estimate of the estimation error.  

\begin{table}[t]
\centering
\caption{\label{tab:results}Estimation results. The values for the
  standard deviation are obtained from simulating and re-estimating
  the respective models 100 times.}%
\fbox{%
\begin{tabular}{lcccc}
estimator & $\alpha$ & $s$ & $\beta$ & $c$ \\\hline
$\hat\vartheta_{\PROJ, \text{ LS}}$ & 
0.296 (0.0193) & 0.234 (0.0744) & 0.379 (0.532) & 1.67 (0.1761)\\ 
$\hat\vartheta_{\SPEC}$ & 
0.338 (0.0166) & 0.687 (0.1797) & 0.456 (0.439) & 2.21 (0.1596)\\ 
$\hat\vartheta_{\SPEC, \text{ CL}}$ & 
0.346 (0.0234) & 1.025 (0.4806) & 0.144 (0.520) & 1.61 (0.1846)
\end{tabular}
}
\end{table}

\iffalse
      [,1]   [,2]   [,3]
[1,] 0.0193 0.0166 0.0234
[2,] 0.0744 0.1797 0.4806
[3,] 0.5403 0.5184 0.5231
[4,] 0.1761 0.1596 0.1846
[1] 0.532 0.439 0.520

options(digits=3)
rbind(
opt0$par,
lambda.stat$lambda.mv$opt.parametricML$par,
lambda.stat$lambda.mv$opt.parametricMLcomp$par,
opt$par[1:2],
lambda.stat$lambda.mv$opt.parametricMLaniso$par[1:2],
lambda.stat$lambda.mv$opt.parametricMLcompAniso$par[1:2])
      [,1]  [,2]
[1,] 0.273 0.156
[2,] 0.324 0.418
[3,] 0.327 0.799
[4,] 0.296 0.234
[5,] 0.338 0.687
[6,] 0.346 1.025
rbind(
opt$par[3:4],
lambda.stat$lambda.mv$opt.parametricMLaniso$par[3:4],
lambda.stat$lambda.mv$opt.parametricMLcompAniso$par[3:4]) %$
     [,1] [,2]
[1,] 0.379 1.67
[2,] 0.456 2.21
[3,] 0.144 1.61
all.par <- NULL
for (i in 1:n.rep){
  all.par <-
   rbind(all.par,
    c(opt.resim[[i]]$par,
      lambda.resimulated[[2]][[i]]$lambda.mv$opt.parametricMLaniso$par,
      lambda.resimulated[[3]][[i]]$lambda.mv$opt.parametricMLcompAniso$par))
} #%$
all.par[,c(3,7,11)] <- angle.trafo(all.par[,c(3,7,11)])
matrix(nr=4, apply(all.par, 2, sd))
apply(all.par[,c(3,7,11)], 2,
         function(x) {
           m <- mean(x)
           sqrt(1/length(x) * sum( pmin( (x-m)^2, (x+pi-m)^2, (x-pi-m)^2) ))
         })
      [,1]   [,2]   [,3]
[1,] 0.0193 0.0166 0.0234
[2,] 0.0744 0.1797 0.4806
[3,] 0.5403 0.5184 0.5231
[4,] 0.1761 0.1596 0.1846
[1] 0.532 0.439 0.520

\fi

\begin{figure}[!htb]
  \caption{\label{fig:realization.ECF.aniso}Level lines of 
    the extremal coefficient
    function and realizations of the fitted Brown-Resnick
    process. Left: Without transformation. Right: After space
    transformation.}
\begin{center}
  \includegraphics[width=0.4\textwidth]{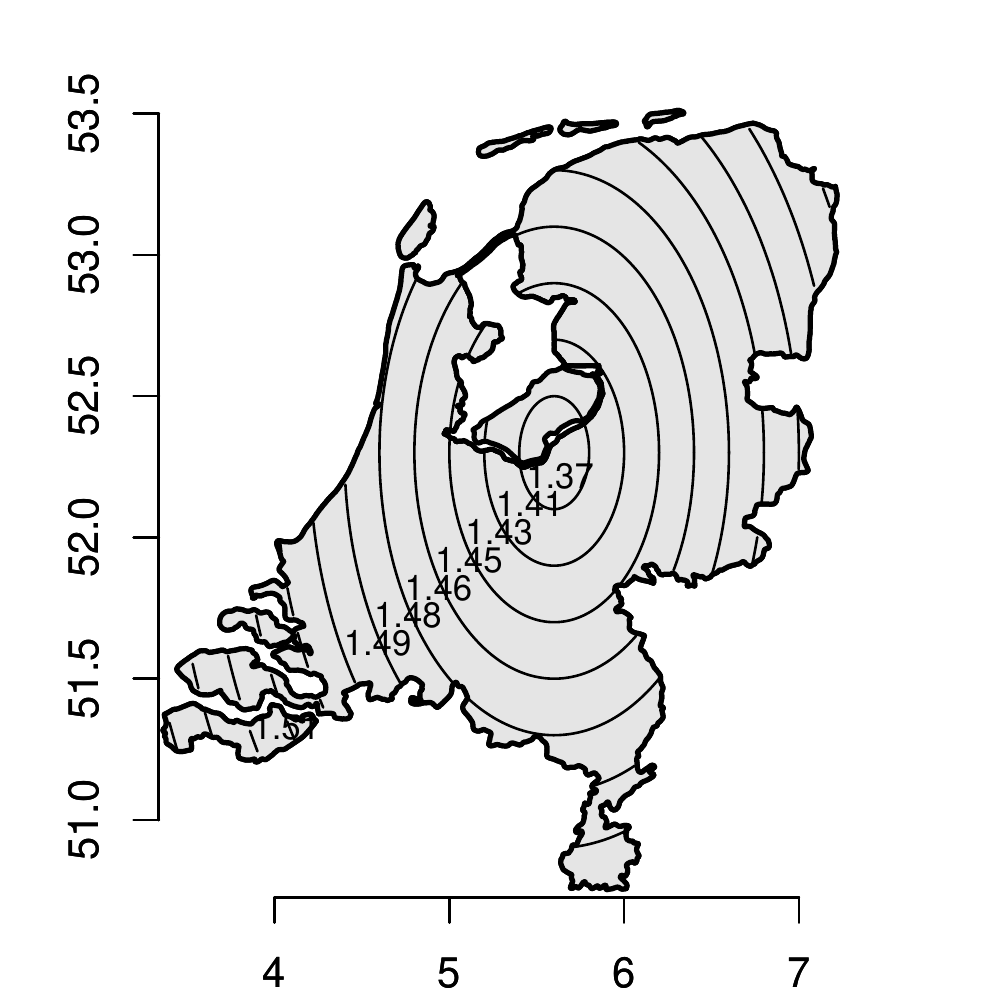}
  \includegraphics[width=0.4\textwidth]{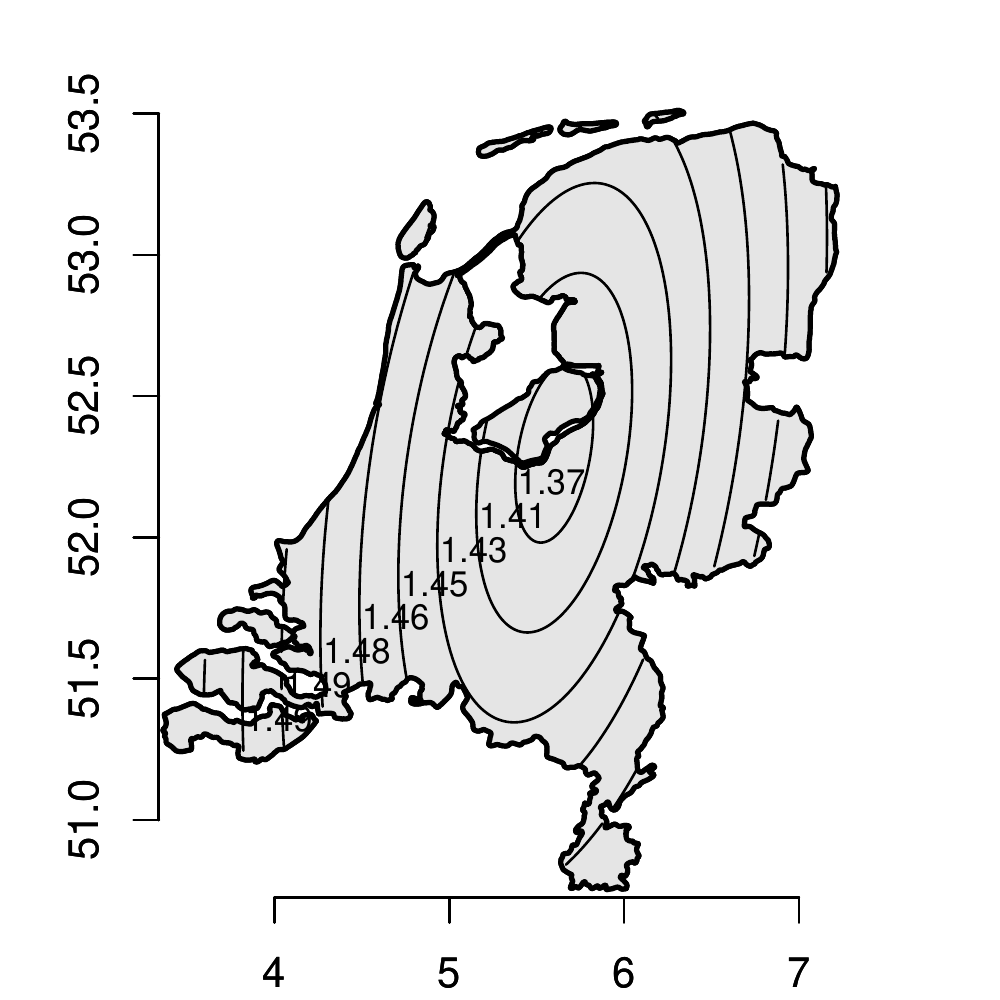}
  \includegraphics[width=0.4\textwidth]{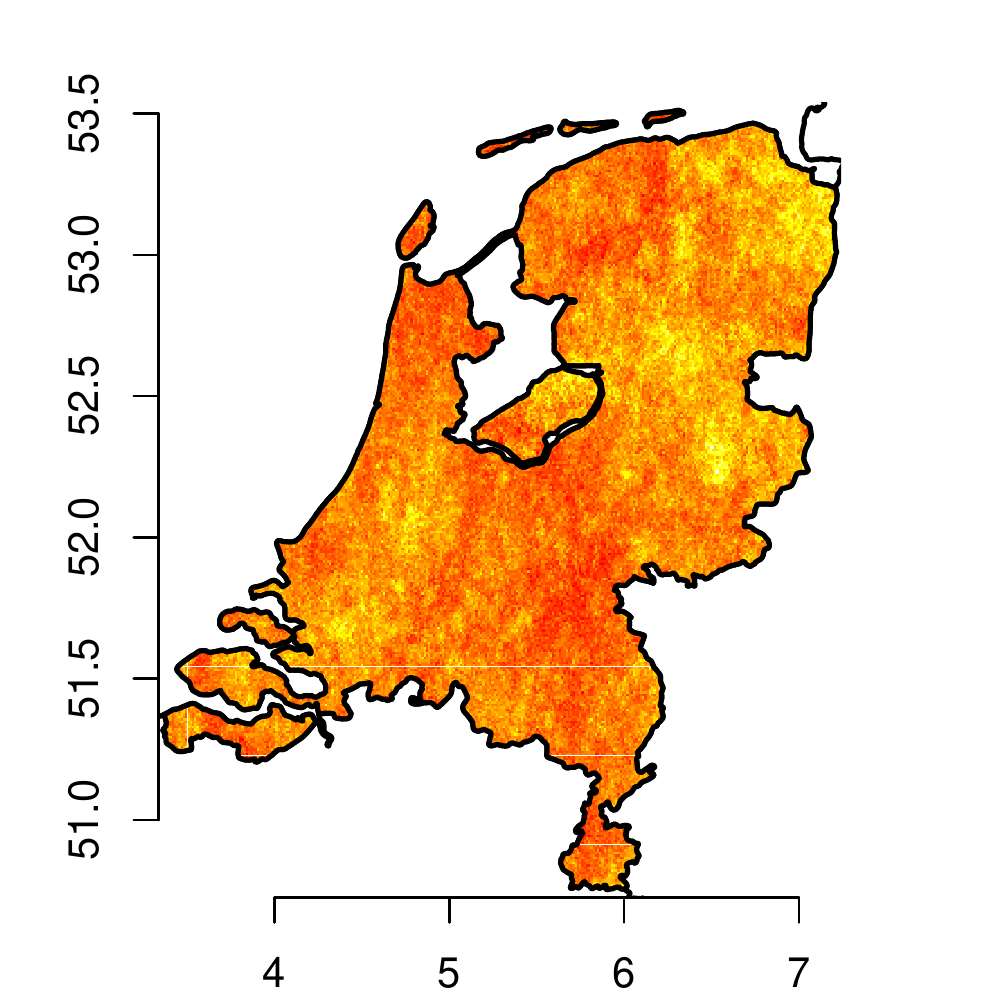}
  \includegraphics[width=0.4\textwidth]{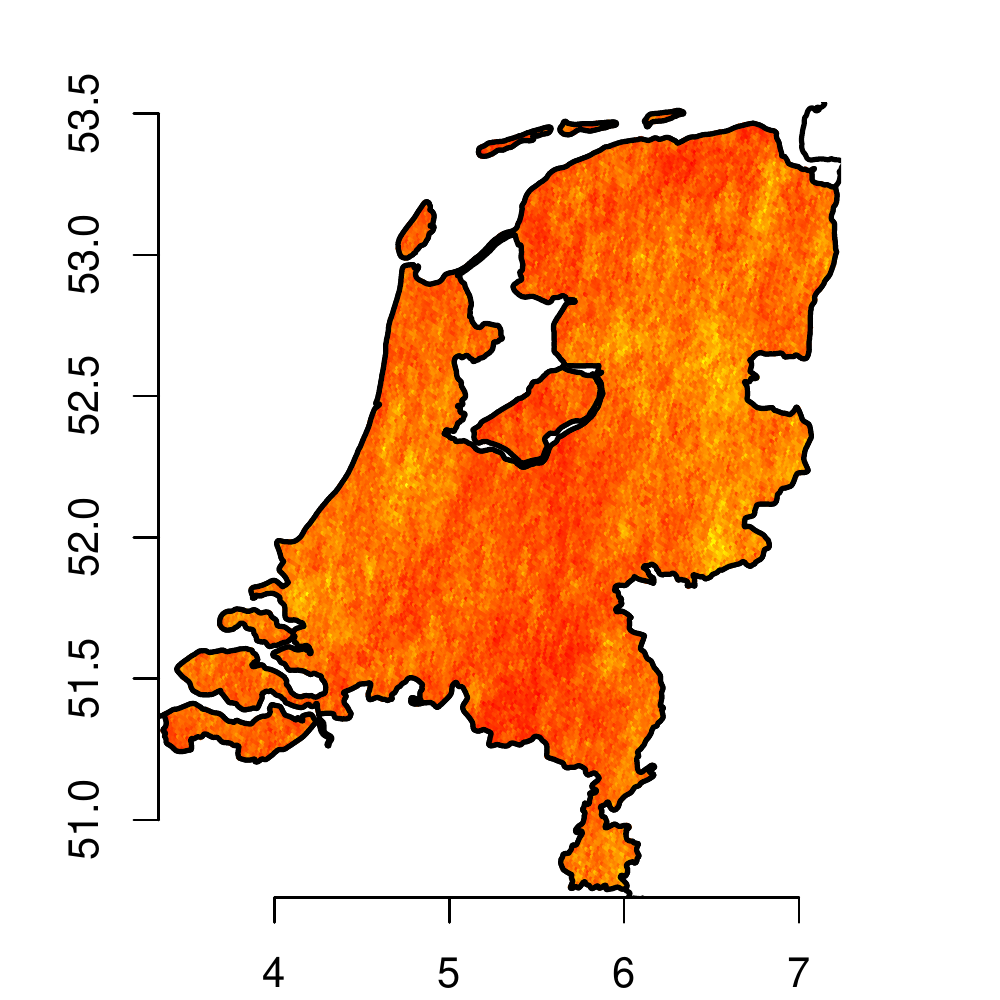}
\end{center}
\end{figure}

The estimation results and standard deviations for the parameters
$(\alpha, s, \beta, c)$ are given in Table \ref{tab:results}. The
middle panel of Figure \ref{fig:fitBR.coast.aniso} illustrates the
effect of transforming the space via the matrix $V$ and displays the
fitted extremal coefficient functions for the three estimators in
\eqref{three_est}.  Moreover, the right panel shows the estimates of
pairwise extremal coefficients based on $\hat\lambda^2_{\MLEa}$ and
the model-independent madogram estimator, where the latter exhibits a
considerably larger variation. In Figure
\ref{fig:realization.ECF.aniso}, we illustrate the effect of
transforming the space via the matrix $V(\beta, c)$ on the extremal
coefficient function and on a typical realization of the corresponding
Brown-Resnick process.

\begin{figure}[!htb]  
  \caption{ \label{fig:fitBR.resim}Validation of estimation: Fitted
    extremal coefficient functions for 100 simulations of 8172 
    Brown-Resnick processes on the
    transformed locations according to the estimated parameters. }
\begin{center}
  \includegraphics[width=0.33\textwidth, page=1, clip=true]{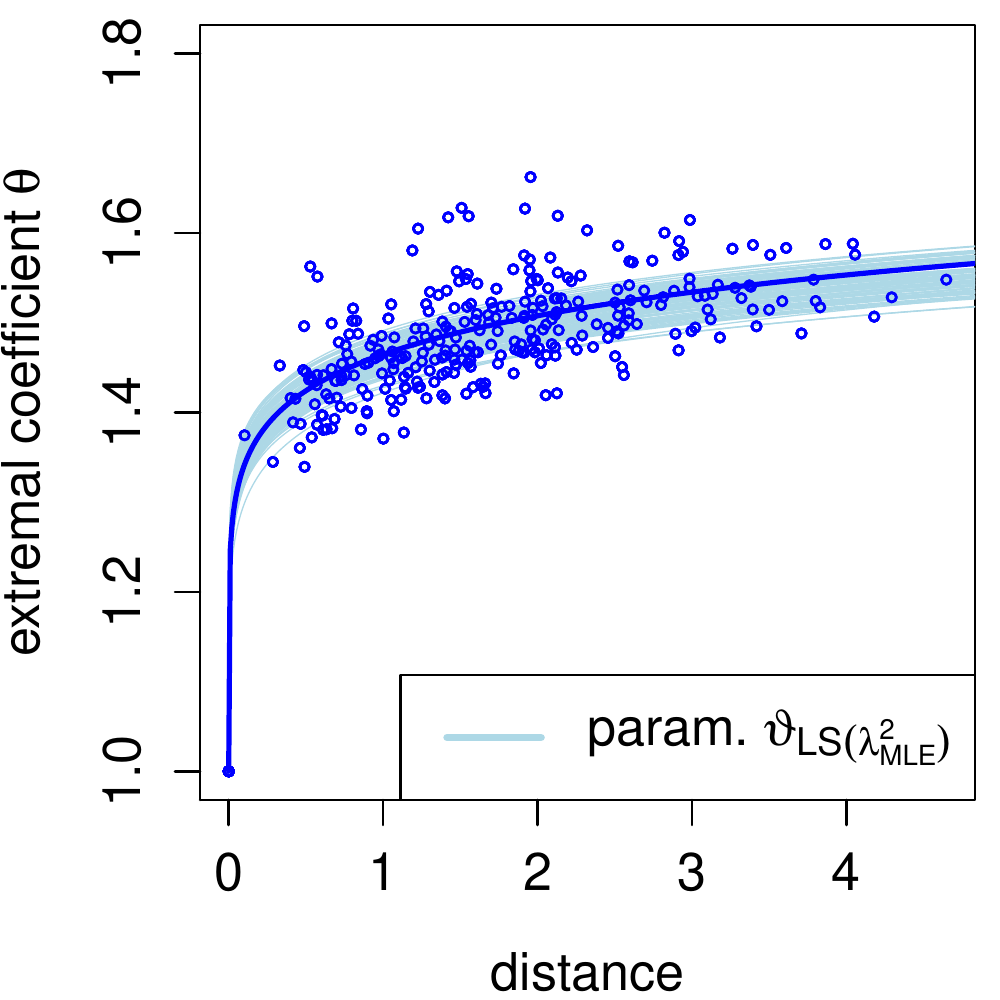}%
  \includegraphics[width=0.33\textwidth, page=2, clip=true]{fig.fitBRvariogram1990.2012.resim}%
  \includegraphics[width=0.33\textwidth, page=3, clip=true]{fig.fitBRvariogram1990.2012.resim}%
\end{center}
\end{figure}

In order to validate the reliability of the estimated model parameters
$\vartheta$, we re-simulate data in the MDA of the three fitted
Brown-Resnick models. Similarly to the simulation study, we use 8172
realizations of the Brown-Resnick process itself (which is clearly in
its own MDA) for the daily data. As index set, we use the transformed
locations $V(\hat{\beta},\hat{c}) T$ on which the Brown-Resnick
process is isotropic. Based on this new data, we apply the estimation
procedure exactly as for the real data to obtain new estimates for
$\vartheta$ and thus for the extremal coefficient function.  This is
repeated 100 times and the results for the three different estimators
in \eqref{three_est} are shown in Figure \ref{fig:fitBR.resim}.  In
agreement with the results of the simulation study, the multivariate
estimator $\hat{\vartheta}_{\SPEC}$ seems to be most reliable
since the re-estimated extremal coefficient functions are close to the
true value of the simulation. In contrast, the composite likelihood
estimator $\hat{\vartheta}_{\SPEC,\text{ CL}}$ significantly
underestimates the true degree of extremal dependence. This is
probably a result of the false assumption of independence of bivariate
densities which underlies the concept of composite likelihoods.

\clearpage

\subsection{Non-stationarity}
\label{non_stat}

In the previous subsection we excluded the $10$ coastal stations from
fitting the stationary Brown-Resnick model since they exhibit a
different extremal dependence structure than the $25$ inland
stations. Here we fit a multivariate H\"usler-Reiss distribution as
extreme value model, which does not rely on any stationarity
assumption.  In particular, for $T'$ being any subset of the 35
locations $x_1,\dots,x_{35} \in \Xcal$, we estimate the $k(k-1)/2$
parameters of the dependence matrix $\Lambda\in\R^{k\times k}$, where
$k = |T'|$.  To this end, we can use any of the three newly proposed
estimators $\hat{\Lambda}_{\Var}$, $\hat{\Lambda}_{\MLEa}$ and
$\hat{\Lambda}_{\SPEC}$. While $\hat{\Lambda}_{\Var}$ is given in
explicit form and hence computationally very efficient and applicable
to arbitrary dimensions, the latter two estimators require numerical
optimization. Fortunately, the respective likelihood functions can be
still evaluated much faster than most of the commonly used spectral
density models. For the ML algorithm, $\hat{\Lambda}_{\Var}$ and,
since the class of H\"usler-Reiss distributions is closed, also the
lower-dimensional parameter estimates provide reasonable starting
values.

\begin{figure}[!htb]
\caption{\label{fig:spectralmodels}Comparison of different spectral density models based on the maximized likelihood. The numbers above the boxes show the average computing time (in seconds) for the numerical maximization.}
\begin{center}
  \includegraphics[width=1\textwidth]{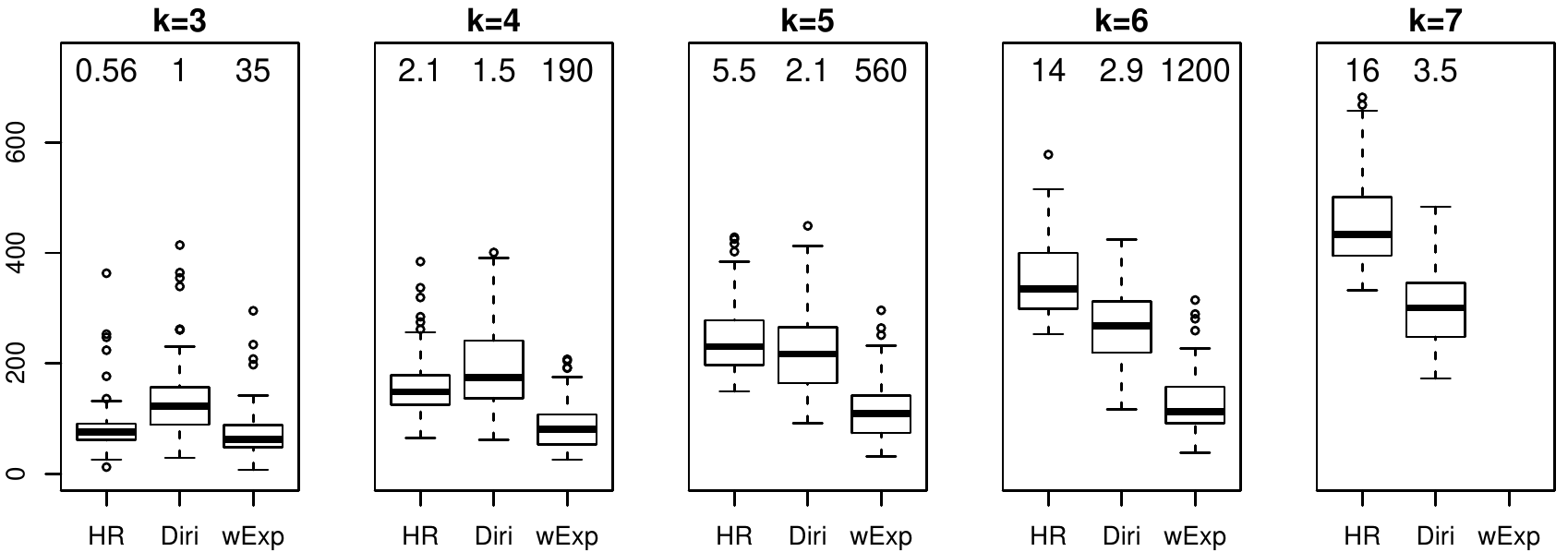}
\end{center}
\end{figure}

In what follows, we use $\hat{\Lambda}_{\Var}$ as a starting value for
the numerical optimization of $\hat{\Lambda}_{\SPEC}$.  We compare the
likelihood values of the H\"usler-Reiss model fit to those of two
other parametric models for spectral densities, namely the Dirichlet
model \citep{col1991} and the weighted exponential model
\citep{bal2011}. The comparison is based on randomly drawing $k=3, 4,
5, 6$ and 7 out of the 35 stations and fitting all three models.  This
is repeated 100 times.  The weighted exponential model seems to fit
worst for all $k\in\{3, 4, 5, 6\}$. Note that numerical optimization
for this model involves a rather complicated likelihood and is
extremely time-consuming. This is why the weighted exponential model
is only included for $k\in\{3, 4, 5, 6\}$. The H\"usler-Reiss model
seems to outperform the Dirichlet model for $k\geq 5$, which is not
completely surprising since the Dirichlet model has only $k$
parameters, while the H\"usler-Reiss model has $k(k-1)/2$ parameters
encoding the extremal dependence. The results are summarized by Figure
\ref{fig:spectralmodels}, which shows boxplots of the maximum
likelihood values for each of the 100 choices of stations, and Table
\ref{tab:spectralmodels}, which shows the percentage of cases in which
the H\"usler-Reiss model outperforms the Dirichlet and the weighted
exponential model.

\iffalse
1] 0.1
[1] 0.79
[1] 99
[1] 0.253
[1] 0.956
[1] 90
[1] 0.733
[1] 0.933
[1] 71
[1] 1
[1] 1
[1] 67
[1] 0.989
[1] NaN
[1] 91
\fi

\begin{table}
\centering
\caption{\label{tab:spectralmodels}Fraction of cases in which the
  H\"usler-Reiss model outperforms the Dirichlet and the weighted exponential model.}
\fbox{
\begin{tabular}{lccccc}
number of stations & $k=3$ & $k=4$ & $k=5$ & $k=6$ & $k=7$   \\[3pt]
$\P(L_{\text{HR}} > L_{\text{Diri}})$ & 0.10 & 0.25 & 0.73 & 1.00 & 0.99\\
$\P(L_{\text{HR}} > L_{\text{wExp}})$ & 0.79 & 0.96 & 0.93 & 1.00 & --\\
\end{tabular}
}
\end{table}

\section{Discussion}
This paper presents several new estimators for the H\"usler-Reiss
distribution and its spatial analog, the Brown-Resnick process. The
methods are based on asymptotic conditional distributions of random
variables in the MDA of the H\"usler-Reiss model. Within the framework
of multivariate peaks-over-threshold, it is shown how conditioning on
different extreme events leads to different estimators. In particular,
the concept of extremal increments turns out to be fruitful, since for
the H\"usler-Reiss model the increments conditioned on a fixed
component being large are approximately multivariate Gaussian
distributed. This enables very efficient inference even for high
dimensions. The simulation study shows, that the proposed estimators
perform well, both in terms of bias and variance. Especially for small
data sets they outperform classical block methods. Moreover, the
non-parametric, bivariate estimators are a suitable tool for
exploratory data analysis (such as distinguishing between coastal and
inland stations in Section \ref{wind_stationary}), since they are
computationally efficient and yield reliable results.\\ With regard to
spatial extreme value statistics, one of the most promising models is
the class of Brown-Resnick processes, due to their flexibility in
connection with parametric families of variograms. The paper provides
several methods for parametric fitting of these models. Particularly
the good performance of the multivariate spectral estimator suggests
that using higher-dimensional densities better captures the shape of
the underlying variogram than methods based on bivariate distributions
only. Also for multivariate analysis of non-stationary data, the
H\"usler-Reiss model is shown to be both well fitting and applicable
in high dimensions due to low computational costs of the
estimators (Section~\ref{non_stat}).\\ While the simulation
study in Section \ref{sim_study} 
already provides some empirical evidence for the consistency of
the proposed estimators, a deeper analysis of the theoretical
properties such as speed of convergence is left for future research.
The main difficulty is to find appropriate assumptions such that the
conditional increments converge not only in distribution but also in
$L_1$ or $L_2$.\\ The idea of including all single extreme events into
statistical inference, in connection with the concept of conditional
increments, might also be applicable to other max-stable models such
as mixed moving maxima processes.

\section*{Supplementary material}
The raw data for the application can be downloaded from http://www.knmi.nl.

\section*{Acknowledgment}
This research has been financially supported by Deutsche Telekom
Stiftung (S. Engelke), by the German Science Foundation
(DFG), Research Training Group 1644 'Scaling problems in Statistics'
(A. Malinowski) and by Volkswagen Stiftung within the
project `WEX-MOP' (M. Schlather).

\section*{Appendix}

\begin{proof}[of Theorem \ref{thm_conv}]
  Note that
  \begin{align*}
    \P\left\{ g(\tilde{\Xbf}_n) \in B \,\big|\, 
    \tilde{\Xbf}_n \in A - \log u(n) \right\} &= 
    \P\left\{ g(\tilde{\Xbf}_n+\log u(n)) \in B \,\big|\, 
    \tilde{\Xbf}_n \in A - \log u(n) \right\} \\
  &= \frac{\P\left\{ \tilde{\Xbf}_n 
      \in (g^{-1}(B)-\log u(n))\cap (A - \log u(n)) \right\}}
          {\P\left\{ \tilde{\Xbf}_n \in A - \log u(n) \right\}}\\
          & = \frac{n/u(n)\P\left\{ \Xbf - \log (n/u(n))
            \in g^{-1}(B)\cap A \right\}}
          {n/u(n)\P\left\{ \Xbf - \log (n/u(n))  \in A \right\}}.
  \end{align*}
  Thus, applying
  Prop. 5.17 in \citet{res2008}, we obtain
  \begin{align*}   
    \lim_{n\to\infty }\P\left\{ g(\tilde{\Xbf}_n) \in B \,\big|\, 
    \tilde{\Xbf}_n \in A - \log u(n) \right\}
        = \frac{ \mu(g^{-1}(B)\cap A)}{\mu(A)}.
  \end{align*}
  and the measure $Q_{g,A}$ is given by 
  \begin{align*}   
    Q_{g,A}(B) = \frac{ \mu(g^{-1}(B)\cap A)}{\mu(A)}, 
    \quad  B\in\mathcal{B}(S).
  \end{align*}

\end{proof}

\begin{proof}[of Theorem \ref{thm_HR}]
  \begin{enumerate}
  \item
    The density of the exponent measure $\mu$ of the H\"usler-Reiss distribution
    \eqref{HRdistr} is given by
    \begin{align*}
      \mu(dx_0,\dots,dx_k) = \frac{e^{-x_0}}{(2\pi)^{\frac{k}{2}}|\text{det} \Sigma|^{1/2}}
      \exp\left(-\frac12 \mathbf{y}^\top\Sigma^{-1}\mathbf{y}
      \right)dx_0\dots dx_k,\quad x_0,\dots,x_k \in\R,
    \end{align*}
    where $\mathbf{y} = (x_1 - x_0 + 2\lambda^2_{1,0},\dots,x_k - x_0 +
    2\lambda^2_{k,0})^\top$ and $\Sigma = \Psi_{k,(0,\dots,k)}(\Lambda)$.
    For $\mathbf{s} = (s_1,\dots,s_k)\in\R^{k}$,
    $B_{\mathbf{s}} = \{\mathbf{y}\in\R^k: y_i \leq s_i, i=1,\dots,k\}$, note
    that $g^{-1}(B_\mathbf{s}) = \{\mathbf{x}\in\R^{k+1}: x_i - x_0 \leq s_i, i=1,\dots,k\}$.
    Thus, for $A_1 = (0,\infty) \times \R^k$, we obtain
    \begin{align*}
      \mu(g^{-1}(B_{\mathbf{s}})\cap A_1) &= \int_0^\infty e^{-x_0}
      \int_{-\infty}^{x_0+s_1}\hdots\int_{-\infty}^{x_0+s_k}
      \frac{\exp\left(-\frac12 \mathbf{y}^\top\Sigma^{-1}\mathbf{y}
      \right)}{(2\pi)^{\frac{k}{2}}|\text{det} \Sigma|^{1/2}}
       dx_k\dots dx_0\\ & =
      \Phi_{M,\Sigma}(s_1,\dots,s_k),
    \end{align*}
    where $\Phi_{M,\Sigma}$ is the cumulative distribution
    function of a $k$-variate normal distribution with mean $M =
    (-2\lambda^2_{1,0}, \dots, -2\lambda^2_{k,0})^\top$ and covariance matrix
    $\Sigma = \Psi_{k,(0,\dots,k)}(\Lambda)$. Since $\mu(A_1) = 1$ and the 
    family of sets $\{B_\mathbf{s},\mathbf{s}\in\R^k\}$ is a generator of the 
    Borel $\sigma$-algebra $\mathcal{B}(\R^k)$ on $\R^k$, the first
    assertion follows from the proof of Theorem \ref{thm_conv}.
    
  \item
    In the bivariate case the density of $\mu$ simplifies to
    \begin{align*}
      \mu(dx,dy) = \frac{e^{-x}}{2\lambda}\phi\left(\lambda +
      \frac{y-x}{2\lambda}\right)dx\,dy, \quad x,y\in\R,
    \end{align*}
    with $\lambda = \lambda_{0,1}$.
    We consider the set $A_2 = [-\infty, \mathbf{0}]^C\subset \R^2$ 
    and note that $\mu(A_2) = 2\Phi(\lambda)$. It thus suffices 
    to compute $\mu(g^{-1}(B_t)\cap A_2)$ for $t\in\R$. For $t<0$ we have
    \begin{align*}
      \mu(g^{-1}(B_t)\cap A_2) = \int_0^\infty \int_{-\infty}^{x+t}
      \frac{e^{-x}}{2\lambda}\phi\left(\lambda +
      \frac{y-x}{2\lambda}\right) dy \, dx = \Phi\left(\lambda +
      \frac{t}{2\lambda}\right),
    \end{align*}
    and similarly for $t>0$,
    \begin{align*}
      \mu(g^{-1}(B_t)\cap A_2) = \mu(A_2) - \Phi\left(\lambda - \frac{t}{2\lambda}\right).
    \end{align*}
    By the above considerations and the proof of Theorem \ref{thm_conv} this yields
    \begin{align*}
      \lim_{n\to\infty} &\P\left\{ X^{(1)} -  X^{(0)} \leq t
      \Big| \tilde{X}_n^{(0)} > \log u(n) \text{ or } \tilde{X}_n^{(1)} > \log u(n)\right\}\\ & =
      \begin{cases}
        \frac{1}{2\Phi(\lambda)}\Phi\left(\lambda +
        \frac{t}{2\lambda}\right) & \text{for } t<0,\\ 1 -
        \frac{1}{2\Phi(\lambda)}\Phi\left(\lambda -
        \frac{t}{2\lambda}\right)& \text{for } t>0.
      \end{cases}
    \end{align*}
    In other words, $X^{(1)} - X^{(0)}$ conditional on either $\tilde{X}_n^{(0)}$
    or $\tilde{X}_n^{(1)}$ being large converges in distribution to some
    random variable $Z$ with density
    \begin{align*}
      g_\lambda(t) = \frac{1}{4\lambda\Phi(\lambda)}\phi\left(\lambda -
      \frac{|t|}{2\lambda}\right), \quad t\in\R.
    \end{align*}
  \end{enumerate}
\end{proof}

\begin{proof}[of Proposition \ref{prop_spec}]
By Theorem 1 in \citet{col1991} we can compute the spectral density
$h$ as a derivative of the exponent measure $\nu(\mathbf{x}) = -\log
G_\Lambda(\mathbf{x})$, namely
\begin{align*}
  \frac{\partial \nu(\mathbf{x})}{\partial x_0\dots\partial x_k} = -
  \left(\sum_{i=0}^k x_i\right)^{-k}
  h\left(\frac{x_0}{\sum_{i=0}^k x_i},\dots,\frac{x_k}{\sum_{i=0}^k
    x_i}; \Lambda \right).
\end{align*}
Since all but one summands of the exponent measure $\nu$ vanish, it
suffices to evaluate
\begin{align}
  \frac{\partial }{\partial x_0\dots\partial x_k} (-1)^k\int_{\log x_0} \int_{\log x_1 -z + 2\lambda^2_{1,0}} \dots \int_{\log x_{k} -z + 2\lambda^2_{k,0}} \phi\left(\mathbf{y} | \Sigma\right) dy_1\dots dy_k \, e^{-z} \, dz,
\end{align}
where $\phi\left(\, \cdot \,| \Sigma\right)$ is the density
function of the $k$-dimensional normal distribution with
covariance matrix $\Sigma$. Carrying out this computation
yields \eqref{spec_den}.
\end{proof}

\bibliography{HREstimation}{}

\begin{thebibliography}{}

\bibitem[\protect\citeauthoryear{Bacro and Gaetan}{Bacro and
  Gaetan}{2012}]{bac2012}
Bacro, J.-N. and C.~Gaetan (2012).
\newblock Estimation of spatial max-stable models using threshold exceedances.
\newblock (Available from \texttt{http://arxiv.org/abs/1205.1107}).

\bibitem[\protect\citeauthoryear{Ballani and Schlather}{Ballani and
  Schlather}{2011}]{bal2011}
Ballani, F. and M.~Schlather (2011).
\newblock A construction principle for multivariate extreme value
  distributions.
\newblock {\em Biometrika\/}~{\em 98}, 633--645.

\bibitem[\protect\citeauthoryear{Blanchet and Davison}{Blanchet and
  Davison}{2011}]{bla2011}
Blanchet, J. and A.~C. Davison (2011).
\newblock Spatial modeling of extreme snow depth.
\newblock {\em Ann. Appl. Stat.\/}~{\em 5}, 1699--1725.

\bibitem[\protect\citeauthoryear{Boldi and Davison}{Boldi and
  Davison}{2007}]{bol2007}
Boldi, M.-O. and A.~C. Davison (2007).
\newblock A mixture model for multivariate extremes.
\newblock {\em J. R. Stat. Soc. Ser. B Stat. Methodol.\/}~{\em 69}, 217--229.

\bibitem[\protect\citeauthoryear{Brown and Resnick}{Brown and
  Resnick}{1977}]{bro1977}
Brown, B.~M. and S.~I. Resnick (1977).
\newblock Extreme values of independent stochastic processes.
\newblock {\em J. Appl. Probab.\/}~{\em 14}, 732--739.

\bibitem[\protect\citeauthoryear{Coles and Tawn}{Coles and
  Tawn}{1991}]{col1991}
Coles, S.~G. and J.~A. Tawn (1991).
\newblock Modelling extreme multivariate events.
\newblock {\em J. R. Stat. Soc. Ser. B Stat. Methodol.\/}~{\em 53}, 377--392.

\bibitem[\protect\citeauthoryear{Cooley, Davis, and Naveau}{Cooley
  et~al.}{2010}]{coo2010}
Cooley, D., R.~A. Davis, and P.~Naveau (2010).
\newblock The pairwise beta distribution: a flexible parametric multivariate
  model for extremes.
\newblock {\em J. Multivariate Anal.\/}~{\em 101}, 2103--2117.

\bibitem[\protect\citeauthoryear{Cooley, Naveau, and Poncet}{Cooley
  et~al.}{2006}]{coo2006}
Cooley, D., P.~Naveau, and P.~Poncet (2006).
\newblock {Variograms for spatial max-stable random fields}.
\newblock In P.~Bertail, P.~Soulier, and P.~Doukhan (Eds.), {\em Dependence in
  Probability and Statistics}, Volume 187 of {\em Lecture Notes in Statistics},
  Chapter~17, pp.\  373--390. New York: Springer.

\bibitem[\protect\citeauthoryear{Davison and Gholamrezaee}{Davison and
  Gholamrezaee}{2012}]{dav2012}
Davison, A.~C. and M.~M. Gholamrezaee (2012).
\newblock Geostatistics of extremes.
\newblock {\em Proc. R. Soc. A\/}~{\em 468}, 581--608.

\bibitem[\protect\citeauthoryear{Davison, Padoan, and Ribatet}{Davison
  et~al.}{2012}]{dav2012b}
Davison, A.~C., S.~A. Padoan, and M.~Ribatet (2012).
\newblock Statistical modeling of spatial extremes.
\newblock {\em Statist. Sci.\/}~{\em 27}, 161--186.

\bibitem[\protect\citeauthoryear{Davison and Smith}{Davison and
  Smith}{1990}]{dav1990}
Davison, A.~C. and R.~L. Smith (1990).
\newblock Models for exceedances over high thresholds.
\newblock {\em J. R. Stat. Soc. Ser. B Stat. Methodol.\/}~{\em 52}, 393--442.

\bibitem[\protect\citeauthoryear{de~Haan and Ferreira}{de~Haan and
  Ferreira}{2006}]{deh2006a}
de~Haan, L. and A.~Ferreira (2006).
\newblock {\em Extreme value theory}.
\newblock Springer Series in Operations Research and Financial Engineering. New
  York: Springer.

\bibitem[\protect\citeauthoryear{de~Haan and Pereira}{de~Haan and
  Pereira}{2006}]{deh2006}
de~Haan, L. and T.~T. Pereira (2006).
\newblock Spatial extremes: models for the stationary case.
\newblock {\em Ann. Statist.\/}~{\em 34}, 146--168.

\bibitem[\protect\citeauthoryear{Dombry, {\'E}yi-Minko, and Ribatet}{Dombry
  et~al.}{2011}]{dom2011}
Dombry, C., F.~{\'E}yi-Minko, and M.~Ribatet (2011).
\newblock Conditional simulations of {B}rown-{R}esnick processes.
\newblock (Available from \texttt{http://arxiv.org/abs/1112.3891}).

\bibitem[\protect\citeauthoryear{Engelke, Malinowski, Oesting, and
  Schlather}{Engelke et~al.}{2012}]{eng2012}
Engelke, S., A.~Malinowski, M.~Oesting, and M.~Schlather (2012).
\newblock Representations of max-stable processes based on single extreme
  events.
\newblock (Available from \texttt{http://arxiv.org/abs/1209.2303}).

\bibitem[\protect\citeauthoryear{Genton, Ma, and Sang}{Genton
  et~al.}{2011}]{gen2011}
Genton, M.~G., Y.~Ma, and H.~Sang (2011).
\newblock On the likelihood function of {G}aussian max-stable processes.
\newblock {\em Biometrika\/}~{\em 98}, 481--488.

\bibitem[\protect\citeauthoryear{Hashorva}{Hashorva}{2006}]{has2006}
Hashorva, E. (2006).
\newblock On the multivariate {H}\"usler-{R}eiss distribution attracting the
  maxima of elliptical triangular arrays.
\newblock {\em Statist. Probab. Lett.\/}~{\em 76}, 2027--2035.

\bibitem[\protect\citeauthoryear{Hashorva, Kabluchko, and W{\"u}bker}{Hashorva
  et~al.}{2012}]{has2012}
Hashorva, E., Z.~Kabluchko, and A.~W{\"u}bker (2012).
\newblock Extremes of independent chi-square random vectors.
\newblock {\em Extremes\/}~{\em 15}, 35--42.

\bibitem[\protect\citeauthoryear{H{\"u}sler and Reiss}{H{\"u}sler and
  Reiss}{1989}]{hue1989}
H{\"u}sler, J. and R.-D. Reiss (1989).
\newblock Maxima of normal random vectors: between independence and complete
  dependence.
\newblock {\em Statist. Probab. Lett.\/}~{\em 7}, 283--286.

\bibitem[\protect\citeauthoryear{Joe}{Joe}{1990}]{joe1990}
Joe, H. (1990).
\newblock Families of min-stable multivariate exponential and multivariate
  extreme value distributions.
\newblock {\em Statist. Probab. Lett.\/}~{\em 9}, 75--81.

\bibitem[\protect\citeauthoryear{Kabluchko}{Kabluchko}{2011}]{kab2011}
Kabluchko, Z. (2011).
\newblock Extremes of independent {G}aussian processes.
\newblock {\em Extremes\/}~{\em 14}, 285--310.

\bibitem[\protect\citeauthoryear{Kabluchko, Schlather, and de~Haan}{Kabluchko
  et~al.}{2009}]{kab2009}
Kabluchko, Z., M.~Schlather, and L.~de~Haan (2009).
\newblock Stationary max-stable fields associated to negative definite
  functions.
\newblock {\em Ann. Probab.\/}~{\em 37}, 2042--2065.

\bibitem[\protect\citeauthoryear{Kotz and Nadarajah}{Kotz and
  Nadarajah}{2000}]{kot2000}
Kotz, S. and S.~Nadarajah (2000).
\newblock {\em Extreme value distributions}.
\newblock London: Imperial College Press.
\newblock Theory and applications.

\bibitem[\protect\citeauthoryear{Oesting, Kabluchko, and Schlather}{Oesting
  et~al.}{2012}]{oes2012}
Oesting, M., Z.~Kabluchko, and M.~Schlather (2012).
\newblock Simulation of {B}rown-{R}esnick processes.
\newblock {\em Extremes\/}~{\em 15}, 89--107.

\bibitem[\protect\citeauthoryear{Padoan, Ribatet, and Sisson}{Padoan
  et~al.}{2010}]{pad2010}
Padoan, S.~A., M.~Ribatet, and S.~A. Sisson (2010).
\newblock Likelihood-based inference for max-stable processes.
\newblock {\em J. Amer. Statist. Assoc.\/}~{\em 105}, 263--277.

\bibitem[\protect\citeauthoryear{Resnick}{Resnick}{2008}]{res2008}
Resnick, S.~I. (2008).
\newblock {\em Extreme values, regular variation and point processes}.
\newblock Springer Series in Operations Research and Financial Engineering. New
  York: Springer.

\bibitem[\protect\citeauthoryear{Rootz{\'e}n and Tajvidi}{Rootz{\'e}n and
  Tajvidi}{2006}]{roo2006}
Rootz{\'e}n, H. and N.~Tajvidi (2006).
\newblock Multivariate generalized {P}areto distributions.
\newblock {\em Bernoulli\/}~{\em 12}, 917--930.

\bibitem[\protect\citeauthoryear{Schlather}{Schlather}{2002}]{sch2002}
Schlather, M. (2002).
\newblock Models for stationary max-stable random fields.
\newblock {\em Extremes\/}~{\em 5}, 33--44.

\bibitem[\protect\citeauthoryear{Schlather and Tawn}{Schlather and
  Tawn}{2003}]{sch2003}
Schlather, M. and J.~A. Tawn (2003).
\newblock A dependence measure for multivariate and spatial extreme values:
  Properties and inference.
\newblock {\em Biometrika\/}~{\em 90}, 139--156.

\bibitem[\protect\citeauthoryear{Smith}{Smith}{1990}]{smi1990}
Smith, R. (1990).
\newblock Max-stable processes and spatial extremes.
\newblock {\em Unpublished manuscript\/}.

\bibitem[\protect\citeauthoryear{Tawn}{Tawn}{1990}]{taw1990}
Tawn, J.~A. (1990).
\newblock Modelling multivariate extreme value distributions.
\newblock {\em Biometrika\/}~{\em 77}, 245--253.

\end{thebibliography}
\bibliographystyle{Chicago}

\end{document}